\documentclass[aps,10pt,prd,notitlepage,showpacs,nofootinbib,superscriptaddress]{revtex4-2}
\usepackage{graphicx}
\usepackage[utf8]{inputenc} 
\usepackage{amsmath}
\usepackage{amssymb}
\usepackage{siunitx}
\usepackage[section]{placeins}
\usepackage{float}
\usepackage{comment}
\usepackage{slashed}
\usepackage[normalem]{ulem}
\usepackage{braket}
\usepackage{comment}

\usepackage{caption}
\usepackage{subcaption}
\usepackage[usenames,dvipsnames]{color}
\usepackage[colorinlistoftodos]{todonotes}
\usepackage[colorlinks=true,citecolor=blue,urlcolor=magenta, pdfborder={0 0 0}]{hyperref}

\usepackage{soul}

\usepackage[commandnameprefix=always]{changes}

\usepackage{slashed}

\usepackage{changes}
\begin{document}
\title{Dilution of cosmological relics and gravitational waves from a supercooled electroweak phase transition in classically scale-invariant models}

\author{Arnab Chaudhuri}
\email{arnab.chaudhuri@vit.ac.in}
\affiliation{School of Advanced Sciences,
Vellore Institute of Technology, Vellore, Tamil Nadu 632014, India.}

\begin{abstract}
In classically scale-invariant extensions of the Standard Model the electroweak scale is generated radiatively, and the absence of tree-level mass terms keeps the symmetric phase metastable down to very low temperatures. We study the resulting phase transition in a model with a real singlet and several additional scalars coupled to the Higgs doublet and to the singlet, in which electroweak symmetry is broken along a Gildener-Weinberg flat direction and a light dilaton appears in the spectrum. Computing the bounce along the two-dimensional tunnelling path and following the false-vacuum fraction through percolation and completion, we find that the transition proceeds in a single, strongly supercooled step and completes above the QCD scale over a region of parameter space consistent with Higgs measurements and perturbativity. Despite the strong supercooling the transition is fast. The bubble walls reach their terminal velocity early, and almost all of the released vacuum energy goes into bulk motion of the plasma, which thermalises and reheats the universe close to the scale set by the vacuum energy. The resulting entropy injection dilutes pre-existing relics by up to several orders of magnitude. It raises the baryon asymmetry that must be present before the transition, relaxes the unitarity bound on the mass of thermal dark matter, and suppresses earlier gravitational-wave backgrounds. If the acoustic description of gravitational-wave production extends to this regime, the signal peaks in the millihertz band, follows an approximate scaling relation between peak amplitude and frequency set by the vacuum energy, and would be observed by LISA throughout the viable region. The results depend on the multiplicity of the new scalars and their Higgs portal coupling at leading order through a single combination, and require the singlet scale to lie at or above the TeV scale.
\end{abstract}

\maketitle
 
%======================================================================
\section{Introduction}
\label{sec:intro}
%======================================================================

The Higgs mass parameter is the only dimensionful coupling of the Standard Model, and its origin remains open. An economical possibility is that the classical Lagrangian contains no mass scale at all and the electroweak scale is generated by quantum corrections, through dimensional transmutation in the effective potential~\cite{Coleman:1973jx}. With several scalars the radiative minimum lies along a flat direction of the tree-level potential, as shown by Gildener and Weinberg~\cite{Gildener:1976ih}, and the pseudo-Goldstone boson of broken scale invariance, the dilaton, is parametrically lighter than the other states. In the Standard Model alone this mechanism fails because the top quark drives the effective potential negative. Extensions with additional scalars or gauge fields can restore it and have been studied extensively~\cite{Bardeen:1995kv,Hempfling:1996ht,Meissner:2006zh,Foot:2007iy,Iso:2009ss,Hambye:2013dgv,Hashino:2015nxa,Marzola:2017jzl,Ahriche:2018rao,Ahriche:2023jdq}. Their most direct collider signatures are a light scalar and a modified Higgs sector.

The same feature that makes these models attractive at zero temperature shapes their cosmology. Without tree-level mass terms the symmetric phase remains a local minimum of the thermal effective potential down to very low temperatures, the barrier to the broken phase is generated by thermal effects, and the transition is generically strongly supercooled~\cite{Witten:1980ez,Randall:2006py,Konstandin:2011dr,Iso:2017uuu,vonHarling:2017yew,Baratella:2018pxi}. Strong supercooling releases a vacuum energy that can exceed the energy of the radiation bath by orders of magnitude. The stochastic gravitational-wave background expected from such transitions has been analysed in a wide range of classically conformal and nearly conformal models~\cite{Jinno:2016knw,Hashino:2016rvx,Marzola:2017jzl,Hashino:2018wee,Prokopec:2018tnq,Ellis:2019oqb,Ellis:2020nnr,Lewicki:2020jiv,Kierkla:2022odc,Ahriche:2023jdq,Athron:2023mer}, and it typically falls in the band of LISA~\cite{LISA:2017pwj,Caprini:2019egz}. More generally, a first-order electroweak transition is a well-studied source of a stochastic background~\cite{Kamionkowski:1993fg,Grojean:2006bp,Huber:2008hg,Caprini:2015zlo,Weir:2017wfa,Caprini:2024hue,Athron:2023xlk,Athron:2023rfq}, and the signal has been computed in many extended Higgs sectors, including singlet and multi-singlet extensions~\cite{Kakizaki:2015wua,Huang:2016cjm,Vaskonen:2016yiu,Beniwal:2017eik,Srivastava:2025oer,Chaudhuri:2025cjp,Naik:2026nyz,Borah:2023zsb,Das:2026zuo}, models with temperature-enhanced scatterings~\cite{Chaudhuri:2025rmd} and dark sectors at a different temperature~\cite{Chaudhuri:2026ovn}. Recent work has refined the treatment of the transition itself, including higher-order thermal corrections and dimensional reduction~\cite{Croon:2020cgk,Kierkla:2023von,Kierkla:2025qyz}, the energy budget of runaway and terminal-velocity walls~\cite{Bodeker:2009qy,Bodeker:2017cim,Ellis:2019oqb,Lewicki:2022pdb}, and the expansion history after inefficient reheating~\cite{Gonstal:2025qky}.

Two questions raised by strong supercooling deserve more attention than they usually receive. The first is whether the transition completes. In a background dominated by vacuum energy the false vacuum inflates, and the onset of nucleation does not guarantee that bubbles percolate or that the false-vacuum volume eventually shrinks~\cite{Guth:1981uk,Guth:1982pn,Turner:1992tz,Ellis:2018mja,Athron:2023xlk}. Below a temperature of order a GeV the transition can instead be triggered by QCD effects~\cite{Witten:1980ez,Iso:2017uuu,vonHarling:2017yew}, which changes the picture qualitatively. The second question concerns what happens to the released energy once the transition is over. The plasma is reheated close to the scale set by the vacuum energy, while species that decoupled before the transition do not share in this reheating, so the entropy injected by the transition dilutes every pre-existing relic. Entropy production at the electroweak phase transition and the resulting dilution of dark matter and of the baryon asymmetry have been studied in extended Higgs sectors with first-order transitions that are not strongly supercooled~\cite{Chaudhuri:2017icn,Chaudhuri:2021agl,Chaudhuri:2021rwt,Chaudhuri:2021ibc,Chaudhuri:2022sis,Chaudhuri:2024vrd}. In a strongly supercooled transition the same mechanism operates as a short period of thermal inflation~\cite{Lyth:1995ka}, and dilution by many orders of magnitude has been invoked to open the parameter space of heavy thermal dark matter~\cite{Hambye:2018qjv,Baldes:2020kam,Baldes:2021aph}. How large the dilution is in a concrete, radiatively broken electroweak sector, and how it correlates with the gravitational-wave signal, has not been established.

We study a classically scale-invariant extension of the Standard Model with a real singlet $s$ and $N_Q$ real scalars $Q_i$ that couple to the Higgs doublet and to the singlet through the portals $\alpha_Q$ and $\beta_Q$. This class of models was introduced and analysed in refs.~\cite{Ahriche:2018rao,Ahriche:2023jdq}, and the gravitational-wave signal of its first-order transitions was studied in ref.~\cite{Ahriche:2023jdq}. The present work differs in emphasis. We follow the strongly supercooled transition through percolation and completion in a vacuum-dominated background, determine the energy budget from wall friction, and make the entropy injection and its consequences for pre-existing relics the main subject, with the gravitational-wave signal as its observable counterpart. Electroweak symmetry is broken along a Gildener-Weinberg flat direction, the $Q_i$ provide the positive contribution to the one-loop potential that stabilises the vacuum, and the dilaton has a mass of about $27$~GeV. A small breaking of the symmetry among the $Q_i$ lets them decay before nucleosynthesis, so that they do not leave a relic abundance in conflict with direct detection. We focus on the region where the transition is a single, strongly supercooled step and follow it from nucleation to completion. The bounce action is computed along the two-dimensional tunnelling path, the false-vacuum fraction is tracked with the vacuum energy included in the expansion rate, and the completion of the transition is checked through the shrinking of the false-vacuum volume. We then determine the energy budget from the friction on the bubble walls, compute the reheating temperature and the entropy injection, and derive their consequences for a pre-existing baryon asymmetry, for thermal relics and for earlier gravitational-wave backgrounds. Finally, we compute the acoustic gravitational-wave spectrum and show that its peak follows an approximate scaling relation set by the vacuum energy. We quantify the dependence on the multiplicity $N_Q$ and on the singlet scale $v_S$, as well as the numerical and theoretical uncertainties of the calculation. Throughout we work in equilibrium thermal field theory; temperature-enhanced scatterings~\cite{Chaudhuri:2025rmd}, a hidden sector at a different temperature~\cite{Chaudhuri:2026ovn} or memory effects in the plasma~\cite{Chaudhuri:2025ylu,Chaudhuri:2026aaz} could shift the nucleation temperature and the duration of the transition, and the results here provide a baseline for such extensions.

The paper is organised as follows. Section~\ref{sec:model} introduces the model, its radiative symmetry breaking and the constraints, and defines three benchmark points. Section~\ref{sec:history} describes the finite-temperature potential and classifies the thermal histories. Section~\ref{sec:perc} determines nucleation, percolation and completion, and section~\ref{sec:energy} the energy budget. Section~\ref{sec:entropy} discusses reheating, entropy injection and the dilution of relics, and section~\ref{sec:gw} the gravitational-wave signal. Section~\ref{sec:NQvS} examines the dependence on $N_Q$ and $v_S$, and section~\ref{sec:conclusions} summarises our conclusions. Appendix~\ref{app:numerics} documents the numerical accuracy of the bounce action, and appendix~\ref{app:scheme} the dependence on the thermal resummation.
 
%======================================================================
\section{Classically scale-invariant scalar sector}
\label{sec:model}
%======================================================================

A theory without dimensionful parameters in its classical action cannot break electroweak symmetry at tree level. Radiative corrections can, as Coleman and Weinberg showed for scalar electrodynamics~\cite{Coleman:1973jx}, and the multi-scalar generalisation of Gildener and Weinberg~\cite{Gildener:1976ih} turns this observation into a predictive framework: the vacuum lies along a direction that is flat at tree level, the one-loop potential lifts the degeneracy, and the state along the flat direction (the scalon, or dilaton) acquires a mass suppressed by a loop factor. Within the Standard Model (SM) alone this mechanism fails, because the top-quark loop makes the one-loop potential unbounded along the Higgs direction. Additional bosonic degrees of freedom are needed, and a large body of work has explored which ones~\cite{Bardeen:1995kv,Hempfling:1996ht,Meissner:2006zh,Foot:2007iy,Iso:2009ss,Hambye:2013dgv,Hashino:2015nxa,Marzola:2017jzl,Ahriche:2018rao,Ahriche:2023jdq}. In this section we set up the minimal scalar extension used throughout the paper, fix its renormalisation conditions, and summarise the constraints that delimit the parameter space.

%----------------------------------------------------------------------
\subsection{Field content and tree-level potential}
\label{sec:tree}
%----------------------------------------------------------------------

Besides the SM Higgs doublet $\Phi$, the scalar sector contains a real gauge singlet $S$ and $N_Q$ real gauge singlets $Q_i$ ($i=1,\dots,N_Q$). We impose a $\mathbb{Z}_2$ symmetry under which $S$ and all $Q_i$ change sign together, and an approximate $O(N_Q)$ symmetry acting on the $Q_i$. Classical scale invariance removes all mass terms and cubic couplings. Up to small $O(N_Q)$-breaking terms discussed at the end of this subsection, the renormalisable potential is
\begin{equation}
\begin{aligned}
V_0 ={}& \frac{\lambda_H}{6}\,(\Phi^\dagger\Phi)^2 + \frac{\lambda_S}{24}\,S^4 + \frac{\omega}{2}\,(\Phi^\dagger\Phi)\,S^2 \\
&+ \frac{\alpha_Q}{2}\,(\Phi^\dagger\Phi)\,Q_iQ_i + \frac{\beta_Q}{4}\,S^2\,Q_iQ_i + \frac{\lambda_Q}{24}\,(Q_iQ_i)^2 ,
\end{aligned}
\label{eq:V0}
\end{equation}
with a sum over $i$ implied. In unitary gauge we write $\Phi=(0,H/\sqrt{2})^{T}$ and denote the singlet background by $s$. Along these directions the tree-level potential reads
\begin{equation}
V_0(H,s) = \frac{\lambda_H}{24}\,H^4 + \frac{\lambda_S}{24}\,s^4 + \frac{\omega}{4}\,H^2s^2 ,
\label{eq:V0bg}
\end{equation}
and the fields $Q_i$ carry no background. The quartic $\lambda_Q$ therefore does not enter the zero-temperature one-loop potential along these directions, but it contributes to the thermal mass of the $Q_i$ (appendix~\ref{app:scheme}) and to the renormalisation-group analysis of section~\ref{sec:constraints}. We set $\lambda_Q(m_h)=0$. Running from $m_h$ to the scales $\mu\simeq\pi T_p$--$2\pi T_p$ relevant for the transition generates a small $\lambda_Q$, negative below $m_h$, which shifts the thermal mass squared of the $Q_i$ by between $-22\%$ and $+6\%$ at the benchmarks; appendix~\ref{app:scheme} quantifies the effect on the transition.

The vacuum of interest is $\langle H\rangle=v=246.22$~GeV and $\langle s\rangle=v_S$. A nonzero $v_S$ breaks the $\mathbb{Z}_2$ spontaneously. Bubbles then nucleate toward $+v_S$ and $-v_S$ with equal probability, and domain walls form when they percolate~\cite{Zeldovich:1974uw}. Stable walls would eventually dominate the energy density, and we assume that a small explicit breaking of the $\mathbb{Z}_2$, for instance from Planck-suppressed operators, lifts the degeneracy of the two vacua and makes the walls annihilate early~\cite{Saikawa:2017hiv}. A bias many orders of magnitude below the vacuum energy released in the electroweak transition is sufficient for this and leaves the transition studied here unchanged. Once both fields condense, the $Q_i$ acquire the common mass
\begin{equation}
m_Q^2 = \tfrac{1}{2}\,(\alpha_Q v^2 + \beta_Q v_S^2) .
\label{eq:mQ}
\end{equation}
Throughout the paper the portal couplings are quoted in units of $4\pi$, and $v_S=1$~TeV with $N_Q=12$ serves as the reference choice. Section~\ref{sec:NQvS} shows that the results depend on $N_Q$ and $\alpha_Q$ at leading order through the combination $N_Q\alpha_Q^2$, and only weakly on $v_S$ once $v_S\gtrsim1$~TeV.

With an exact $O(N_Q)$ symmetry the $Q_i$ would be stable. After the transition they return to equilibrium, since $m_Q/T_{\rm reh}\approx7.5$ (section~\ref{sec:entropy}), and freeze out at $T\approx12$~GeV through their large portal couplings. The resulting abundance is $\Omega_Q h^2\approx5\times10^{-4}$ per species, about five percent of the dark matter in total, and the spin-independent scattering rate on nuclei is $N_Q$ times that of a thermal Higgs-portal scalar of the same mass and portal coupling. This region is disfavoured by direct detection~\cite{GAMBIT:2017gge,LZ:2024zvo}. We therefore take the $O(N_Q)$ symmetry to be approximate. The $\mathbb{Z}_2$ allows the dimension-four, scale-invariant couplings
\begin{equation}
V_\kappa = \kappa_i\,Q_i\,S\,(\Phi^\dagger\Phi) ,
\label{eq:kappa}
\end{equation}
together with operators of the type $S^3Q_i$. After symmetry breaking eq.~\eqref{eq:kappa} mixes each $Q_i$ with $h$ at an angle $\theta_i\approx\kappa_i v v_S/(m_Q^2-m_h^2)$ and induces a negligible tadpole. A $Q_i$ of mass $300$~GeV then decays like a heavy Higgs boson with a width suppressed by $\theta_i^2$, mainly into $W$ and $Z$ pairs, with a lifetime of about $10^{-2}~{\rm s}\,(10^{-12}/\kappa_i)^2$. Decay before big-bang nucleosynthesis therefore requires only $\kappa_i\gtrsim10^{-12}$. With exactly universal portals, only the combination $\kappa_iQ_i$ would mix with $h$ and the orthogonal combinations would remain stable. Small non-universal splittings of $\alpha_Q$ and $\beta_Q$ lift this degeneracy, so that every mass eigenstate acquires a component along $\kappa_i$. Couplings $\kappa_i$ of this size modify the effective potential at order $\kappa_i^2$, and at fixed average portal coupling the splittings enter the transition only at second order in their size, so the $Q_i$ are treated as $N_Q$ degenerate states in what follows. Small values of $\kappa_i$ are technically natural, because setting them to zero restores the $O(N_Q)$ symmetry.

%----------------------------------------------------------------------
\subsection{One-loop effective potential and renormalisation conditions}
\label{sec:oneloop}
%----------------------------------------------------------------------

The field-dependent masses of the states that dominate the one-loop potential are
\begin{equation}
m_W^2 = \frac{g^2}{4}H^2,\qquad m_Z^2=\frac{g^2+g'^2}{4}H^2,\qquad m_t^2=\frac{y_t^2}{2}H^2,\qquad m_Q^2(H,s)=\frac{\alpha_Q H^2+\beta_Q s^2}{2}.
\label{eq:masses}
\end{equation}
We work in Landau gauge and in the $\overline{\rm DR}$ scheme (dimensional reduction with modified minimal subtraction), in which the one-loop Coleman-Weinberg contribution is~\cite{Coleman:1973jx,Martin:2001vx}
\begin{equation}
V_1(H,s) = \sum_{i} \frac{n_i}{64\pi^2}\, m_i^4(H,s)\left[\ln\frac{m_i^2(H,s)}{\Lambda^2}-\frac{3}{2}\right],
\label{eq:VCW}
\end{equation}
with $n_W=6$, $n_Z=3$, $n_t=-12$ and $n_Q=N_Q$. In this scheme the constant is $3/2$ for all species, whereas in $\overline{\rm MS}$ the gauge bosons carry $5/6$. The difference is a quartic polynomial in the background fields and is absorbed by the renormalisation conditions below. Loops of the scalars $h$, $\eta$ and the Goldstone modes are not included at zero temperature. Their couplings are small in the region of interest ($|\lambda_H|<0.35$, $|\lambda_S|,|\omega|\lesssim 2\times10^{-2}$), and their thermal counterparts change the transition parameters by at most seven percent (appendix~\ref{app:scheme}).

Because every field-dependent mass in eq.~\eqref{eq:masses} is quadratic in the background fields, a change of the renormalisation scale shifts $V_1$ by a quartic polynomial in $H$ and $s$. Such a shift can be absorbed entirely into the three quartics of eq.~\eqref{eq:V0bg}. We therefore treat $\lambda_H$, $\lambda_S$ and $\omega$ as effective one-loop couplings at $\Lambda=m_h$ (tree-level values plus finite counterterms) and fix them by three physical conditions:
\begin{equation}
\left.\frac{\partial V}{\partial H}\right|_{(v,v_S)}=0,\qquad
\left.\frac{\partial V}{\partial s}\right|_{(v,v_S)}=0,\qquad
m_h = 125.18~{\rm GeV},
\label{eq:rencond}
\end{equation}
where $V=V_0+V_1$ and $m_h$ is one of the two eigenvalues of the scalar mass matrix
\begin{equation}
\mathcal{M}^2=\begin{pmatrix} \partial_H^2 V & \partial_H\partial_s V\\ \partial_H\partial_s V & \partial_s^2 V\end{pmatrix}_{(v,v_S)} .
\label{eq:massmatrix}
\end{equation}
The tadpole conditions are linear in the quartics, and the mass condition then reduces to a single equation for $\omega$. For given $(N_Q,\alpha_Q,\beta_Q,v_S)$ the potential is completely determined, the vacuum sits at $(v,v_S)$ by construction, and the result does not depend on $\Lambda$. The eigenvalues of eq.~\eqref{eq:massmatrix} are curvature masses at zero external momentum, and we identify them with the pole masses; the difference comes from the momentum dependence of the one-loop self-energies, which we do not include. This choice is equivalent to fixing the tadpole and Higgs-mass counterterms on shell, as done in earlier studies of classically scale-invariant models with additional scalars~\cite{Ahriche:2023jdq}.

The mass eigenstates are
\begin{equation}
\begin{pmatrix} h\\ \eta\end{pmatrix} = \begin{pmatrix}\cos\theta & \sin\theta\\ -\sin\theta & \cos\theta\end{pmatrix}\begin{pmatrix} H-v\\ s-v_S\end{pmatrix},
\label{eq:mixing}
\end{equation}
where $h$ is the 125~GeV state and $\sin^2\theta$ measures its singlet admixture.

%----------------------------------------------------------------------
\subsection{Radiative symmetry breaking and the light dilaton}
\label{sec:dilaton}
%----------------------------------------------------------------------

Equation~\eqref{eq:rencond} has two classes of solutions, depending on which eigenvalue of $\mathcal{M}^2$ is identified with $m_h$. When the 125~GeV state is the heavier one, the lighter state $\eta$ is mostly singlet and plays the role of the dilaton of the Gildener-Weinberg construction. Its mass is controlled by the one-loop coefficient along the flat direction,
\begin{equation}
m_\eta^2 \simeq \frac{1}{8\pi^2 v_\phi^2}\left[\,6m_W^4+3m_Z^4+m_h^4+N_Q\,m_Q^4-12\,m_t^4\,\right],\qquad v_\phi^2=v^2+v_S^2 ,
\label{eq:mdilaton}
\end{equation}
which holds up to corrections of order $v^2/v_\phi^2$ and higher loops~\cite{Gildener:1976ih}. The bracket must be positive, and with $m_t^4$ entering twelve times this requires $N_Q m_Q^4\gtrsim 12\,m_t^4$. For $N_Q=12$ the $Q$ states therefore have to be heavier than the top quark, and the benchmark points below have $m_Q\simeq300$~GeV. Equation~\eqref{eq:mdilaton} then gives $m_\eta\approx 31$--$36$~GeV, within twenty percent of the exact eigenvalues listed in table~\ref{tab:bp}. The suppression of $m_\eta$ relative to $v_\phi$ by a loop factor is the property that makes the effective potential nearly flat over a wide range of field values, and it is the origin of the strong supercooling studied in the following sections.

The second class of solutions identifies $m_h$ with the lighter eigenvalue. In that case the Higgs mass itself is generated radiatively, while the mostly singlet state is heavy. Section~\ref{sec:history} shows that this branch does not support a supercooled electroweak transition: the singlet direction breaks first, at temperatures of several hundred GeV, and the subsequent electroweak step is continuous. Unless stated otherwise, all results below refer to the light-dilaton branch.

%----------------------------------------------------------------------
\subsection{Theoretical and experimental constraints}
\label{sec:constraints}
%----------------------------------------------------------------------

\paragraph{Higgs signal strength and $h\to\eta\eta$.}
The mixing in eq.~\eqref{eq:mixing} rescales all couplings of $h$ to SM particles by $\cos\theta$. In addition, the dilaton is lighter than $m_h/2$ throughout the light-dilaton branch, so the decay $h\to\eta\eta$ is open. Its width is
\begin{equation}
\Gamma(h\to\eta\eta) = \frac{\lambda_{h\eta\eta}^2}{32\pi\,m_h}\sqrt{1-\frac{4m_\eta^2}{m_h^2}},\qquad
\lambda_{h\eta\eta}=\left.\frac{\partial^3 V}{\partial h\,\partial\eta^2}\right|_{(v,v_S)},
\label{eq:hee}
\end{equation}
with the trilinear coupling normalised as $\mathcal{L}\supset-\frac12\lambda_{h\eta\eta}\,h\,\eta^2$ and taken from the full one-loop potential. The global signal strength of the 125~GeV state is
\begin{equation}
\mu_{hh} = \cos^2\theta\,\big[1-{\rm BR}(h\to\eta\eta)\big],\qquad
{\rm BR}(h\to\eta\eta)=\frac{\Gamma(h\to\eta\eta)}{\cos^2\theta\,\Gamma_h^{\rm SM}+\Gamma(h\to\eta\eta)},
\label{eq:mu}
\end{equation}
with $\Gamma_h^{\rm SM}=4.1$~MeV~\cite{LHCHiggsCrossSectionWorkingGroup:2016ypw}. We adopt the requirement $\mu_{hh}\geq0.89$, which corresponds approximately to the lower end of the two-standard-deviation ranges of the ATLAS and CMS combinations~\cite{ATLAS:2022vkf,CMS:2022dwd}.

This constraint turns out to be the most restrictive one on the light-dilaton branch. The coupling $\lambda_{h\eta\eta}$ grows with the portal couplings, and over much of the region where the electroweak transition is strongly first order it drives ${\rm BR}(h\to\eta\eta)$ to tens of percent. Viable points have a small trilinear coupling: at the benchmark points $\lambda_{h\eta\eta}\approx -1.2$~GeV, $\Gamma(h\to\eta\eta)\approx0.1$~MeV and ${\rm BR}(h\to\eta\eta)\approx2.5\%$. With couplings inherited from the Higgs through mixing, $\eta$ decays dominantly to $b\bar b$, with ${\rm BR}(\eta\to\mu^+\mu^-)\approx3.5\times10^{-4}$ at $m_\eta\approx27$~GeV. Direct searches for exotic Higgs decays into pairs of light scalars~\cite{Curtin:2013fra} are most sensitive at this mass in the $b\bar b\mu^+\mu^-$ final state, where ATLAS excludes ${\rm BR}(h\to\eta\eta\to b\bar b\mu^+\mu^-)$ above a few times $10^{-4}$~\cite{ATLAS:2021hbr}. The benchmark points predict about $1.5\times10^{-5}$, an order of magnitude below the current sensitivity.

\paragraph{LEP searches for a light scalar.}
The dilaton couples to the $Z$ boson with the strength of a Standard Model Higgs boson reduced by $\sin\theta$, so it could have been produced at LEP through $e^+e^-\to Z\eta$. Since all its couplings arise from mixing, its decays are those of a light Standard Model Higgs boson, and the combined LEP bound on $\xi^2=(g_{HZZ}/g_{HZZ}^{\rm SM})^2$ for Standard-Model-like decays~\cite{LEPWorkingGroupforHiggsbosonsearches:2003ing} applies with $\xi^2=\sin^2\theta$. Between $15$ and $35$~GeV the observed limit lies at $\xi^2\simeq0.015$--$0.022$. The singlet admixture grows as $\beta_Q$ decreases, while the dilaton becomes lighter, so the bound excludes points at small $\beta_Q$ that the Higgs signal strength allows, namely $\beta_Q/4\pi\lesssim0.005$ for $\alpha_Q/4\pi\lesssim0.12$--$0.15$, with the boundary moving to larger $\alpha_Q$ as $\beta_Q$ decreases. The benchmarks have $\sin^2\theta$ equal to $0.58$, $0.38$ and $0.09$ times the limit at their dilaton mass. For $m_\eta<12$~GeV, which occurs only at $v_S=3$~TeV in this work, the LEP~1 searches and the bounds on $\Upsilon$ decays to a light scalar would apply instead, and we have not imposed them.

\paragraph{Perturbativity and the Landau pole.}
We require $|\lambda_H|$, $|\lambda_S|$ and $|\omega|$ to be smaller than $4\pi$ at $\mu=m_h$. The portal coupling $\alpha_Q$ is large in the region of interest, $\alpha_Q/4\pi\approx0.12$--$0.18$, and the more relevant limit comes from its running. At one loop the $\beta$-functions of the quartic couplings are the same in the $\overline{\rm MS}$ and $\overline{\rm DR}$ schemes, and the conversion of the effective couplings at $\Lambda=m_h$ into running couplings is a finite shift of order $g^4/16\pi^2$, negligible next to the portal contributions. The $\beta$-functions follow from the general results of refs.~\cite{Machacek:1984zw,Luo:2002ti}. Writing the quartic part of eq.~\eqref{eq:V0} as $\frac{1}{24}\lambda_{abcd}\,\phi_a\phi_b\phi_c\phi_d$ in terms of the $5+N_Q$ real scalar fields $\phi_a=(\Phi,S,Q_i)$, the scalar contribution reads
\begin{equation}
16\pi^2\,\beta_{\lambda_{abcd}} \supset \lambda_{abef}\lambda_{efcd}+\lambda_{acef}\lambda_{efbd}+\lambda_{adef}\lambda_{efbc} .
\label{eq:betaquartic}
\end{equation}
The gauge and top-Yukawa contributions enter only through the Higgs doublet: the anomalous dimension $16\pi^2\gamma_\Phi = 3y_t^2-\frac94 g^2-\frac34 g'^2$ multiplies each coupling once per external doublet leg, and $\lambda_H$ receives in addition the familiar terms $-6y_t^4+\frac38\left[2g^4+(g^2+g'^2)^2\right]$ (in the normalisation $\lambda_H/6=\lambda$ of the SM quartic $\lambda(\Phi^\dagger\Phi)^2$). Contracting eq.~\eqref{eq:betaquartic} reproduces, for instance, the well-known contribution $N_Q\alpha_Q^2/2$ of the singlets to $16\pi^2\beta_\lambda$.
We evolve the system, together with the SM running of $g$, $g'$, $g_s$ and $y_t$, from $\mu=m_h$, and define the Landau-pole scale $\Lambda_{\rm LP}$ as the scale at which the first of $|\lambda_H|$, $|\lambda_S|$, $|\omega|$, $\alpha_Q$, $\beta_Q$ or $|\lambda_Q|$ reaches $4\pi$. This scale is a proxy for the breakdown of perturbation theory rather than the position of a true pole. The dominant effect is the contribution $N_Q\alpha_Q^2$ to the running of $\lambda_H$, which feeds back into $\alpha_Q$. We require $\Lambda_{\rm LP}>10$~TeV. A one-loop $\overline{\rm MS}$ evolution is the more conservative choice: mass-dependent schemes, in which heavy states decouple below their mass thresholds, push the pole to higher scales~\cite{Ahriche:2023jdq}.

The combined effect of these constraints is shown in figure~\ref{fig:history}. The Higgs signal strength removes the region of larger $\beta_Q$, the LEP bound the corner of small $\beta_Q$ and small $\alpha_Q$, triviality limits $\alpha_Q/4\pi$ to values below about $0.18$ for $N_Q=12$, and the strongly supercooled region lies in the strip between them.

%----------------------------------------------------------------------
\subsection{Benchmark points}
\label{sec:benchmarks}
%----------------------------------------------------------------------

We illustrate the results with the three benchmark points of table~\ref{tab:bp}. They share $\beta_Q/4\pi=0.006$, $N_Q=12$ and $v_S=1$~TeV, and differ only in $\alpha_Q$. Along this line the scalar spectrum hardly changes: $m_\eta\approx27$~GeV, $m_Q=295$--$314$~GeV and ${\rm BR}(h\to\eta\eta)\approx2.5\%$. The differences between BP1, BP2 and BP3 therefore reflect the strength of supercooling alone, which ranges from extreme (BP1) through intermediate (BP2) to mild (BP3). As section~\ref{sec:entropy} shows, the transition parameters depend only weakly on $\beta_Q$ across the allowed region (figure~\ref{fig:Dplane}), so this choice does not restrict the generality of the results.

Two trends along the benchmark line deserve a remark. The effective coupling $\lambda_H(m_h)$ falls from $0.32$ at BP1 to essentially zero at BP3 ($-1.1\times10^{-4}$). In other words, the Higgs quartic at the electroweak scale is increasingly generated by the $Q$ loops as $\alpha_Q$ grows. At the same time the singlet admixture of the 125~GeV state decreases from $\sin^2\theta=0.013$ to $0.002$. All three points satisfy the constraints of section~\ref{sec:constraints} with $\mu_{hh}\geq0.96$ and $\Lambda_{\rm LP}\geq18$~TeV. Because $\lambda_H(m_h)$ is slightly negative at BP3, we have checked that $(v,v_S)$ is the global minimum of the zero-temperature one-loop potential at all three benchmarks, on a grid of 181 rays in the $(H,s)$ quadrant extending to field values of 20~TeV. Along $s=0$ the $Q$ loops keep the potential rising, and at 20~TeV it lies about $2\times10^{16}$~GeV$^4$ above the vacuum.

\begin{table}[t]
\centering
\begin{tabular}{lccc}
\hline\hline
 & BP1 & BP2 & BP3 \\ \hline
$\alpha_Q/4\pi$ & 0.13 & 0.14 & 0.16 \\
$\beta_Q/4\pi$ & 0.006 & 0.006 & 0.006 \\ \hline
$\lambda_H(m_h)$ & $3.18\times10^{-1}$ & $2.22\times10^{-1}$ & $-1.13\times10^{-4}$ \\
$\lambda_S(m_h)$ & $1.62\times10^{-3}$ & $1.42\times10^{-3}$ & $9.92\times10^{-4}$ \\
$\omega(m_h)$ & $-1.48\times10^{-2}$ & $-1.44\times10^{-2}$ & $-1.33\times10^{-2}$ \\ \hline
$m_\eta$ [GeV] & 26.9 & 27.0 & 26.4 \\
$m_Q$ [GeV] & 295 & 302 & 314 \\
$\sin^2\theta$ & 0.013 & 0.008 & 0.002 \\
${\rm BR}(h\to\eta\eta)$ & 0.026 & 0.025 & 0.022 \\
$\mu_{hh}$ & 0.962 & 0.967 & 0.976 \\
$\Lambda_{\rm LP}$ [TeV] & 61 & 38 & 18 \\
\hline\hline
\end{tabular}
\caption{Benchmark points for $N_Q=12$ and $v_S=1$~TeV. The quartic couplings are the effective one-loop values at $\mu=m_h$ fixed by the tadpole conditions and $m_h=125.18$~GeV. $\sin^2\theta$ is the singlet admixture of the 125~GeV state and $\Lambda_{\rm LP}$ the scale at which a quartic reaches $4\pi$ under one-loop running. Scalar masses are curvature masses of the one-loop potential.}
\label{tab:bp}
\end{table}
 
%======================================================================
\section{Thermal history}
\label{sec:history}
%======================================================================

The absence of a tree-level mass term has a direct consequence at finite temperature. Thermal loops give the symmetric point a positive curvature of order $T^2$, and nothing in the zero-temperature potential competes with it near the origin. The symmetric phase therefore stays metastable down to arbitrarily low temperatures, and the universe can leave it only by bubble nucleation~\cite{Witten:1980ez,Randall:2006py,Konstandin:2011dr}. This section describes the finite-temperature potential, classifies the possible thermal histories over the $(\alpha_Q,\beta_Q)$ plane, and examines the field-space structure of the transition at the three benchmarks.

%----------------------------------------------------------------------
\subsection{Finite-temperature effective potential}
\label{sec:VT}
%----------------------------------------------------------------------

The one-loop thermal contribution is~\cite{Dolan:1973qd,Weinberg:1974hy}
\begin{equation}
V_T(H,s;T) = \frac{T^4}{2\pi^2}\left[\sum_{b=W,Z,Q} n_b\, J_B\!\left(\frac{m_b^2}{T^2}\right) + n_t\, J_F\!\left(\frac{m_t^2}{T^2}\right)\right],
\label{eq:VT}
\end{equation}
with the multiplicities of eq.~\eqref{eq:VCW}, $n_t=-12$, and
\begin{equation}
\begin{gathered}
J_B(y)=\int_0^\infty dx\,x^2\ln\!\left(1-e^{-\sqrt{x^2+y}}\right),\\
J_F(y)=\int_0^\infty dx\,x^2\ln\!\left(1+e^{-\sqrt{x^2+y}}\right).
\end{gathered}
\label{eq:JBJF}
\end{equation}
Infrared contributions from the bosonic Matsubara zero modes are resummed following Arnold and Espinosa~\cite{Arnold:1992rz},
\begin{equation}
V_{\rm daisy}(H,s;T) = -\frac{T}{12\pi}\sum_{i} n_i\left[\left(\bar m_i^2\right)^{3/2}-\left(m_i^2\right)^{3/2}\right],
\label{eq:daisy}
\end{equation}
where the sum runs over the longitudinal gauge bosons ($n_{W_L}=2$, $n_{Z_L}=n_{\gamma_L}=1$) and the $N_Q$ singlets $Q_i$. The thermally corrected masses $\bar m_i^2$ are the eigenvalues of the field-dependent mass matrices with the Debye masses added. For the gauge bosons we use the leading-order values $\Pi_W=\tfrac{11}{6}g^2T^2$ and $\Pi_B=\tfrac{11}{6}g'^2T^2$ on the diagonal of the $(W^3,B)$ matrix~\cite{Carrington:1991hz}. For the singlets the one-loop Debye mass follows from eq.~\eqref{eq:V0}. The term $\tfrac{\alpha_Q}{2}(\Phi^\dagger\Phi)Q_iQ_i$ couples each $Q_i$ to the four real components of $\Phi$, the term $\tfrac{\beta_Q}{4}S^2Q_iQ_i$ couples it to $S$, and $\lambda_Q(m_h)=0$, so that
\begin{equation}
\Pi_Q = T^2\left(\frac{\alpha_Q}{6}+\frac{\beta_Q}{24}\right).
\label{eq:PiQ}
\end{equation}
A nonzero $\lambda_Q$ would add $(N_Q+2)\lambda_Q T^2/72$ to $\Pi_Q$; appendix~\ref{app:scheme} examines the value generated by running.
The full potential is $V(H,s;T)=V_0+V_1+V_T+V_{\rm daisy}$. Its field-independent part is irrelevant for tunnelling, and the radiation energy density that enters the expansion rate and the transition strength is computed separately, as $\rho_{\rm rad}=\pi^2g_*T^4/30$ with $g_*=106.75+1+N_Q$ in the symmetric phase. This treats the $Q_i$ as massless; their thermal mass, $m_Q/T\simeq0.53$--$0.58$ at $T_p$, lowers their energy density by about four percent, which changes $g_*$ by $-0.4$ to $-0.5$ and is neglected. Thermal loops of $h$, $\eta$ and the Goldstone modes are omitted, as at zero temperature. The resummation prescription is the main source of theoretical uncertainty, and appendix~\ref{app:scheme} compares eq.~\eqref{eq:daisy} with the Parwani scheme~\cite{Parwani:1991gq} and with the inclusion of scalar loops (see also ref.~\cite{Curtin:2016urg}).

Expanding eqs.~\eqref{eq:VT} and \eqref{eq:daisy} around the origin, where every field-dependent mass vanishes, gives the curvatures
\begin{equation}
\begin{gathered}
\left.\frac{\partial^2 V}{\partial H^2}\right|_{0} = c_H T^2,\qquad c_H=\frac{3g^2+g'^2}{16}+\frac{y_t^2}{4}+\frac{N_Q\alpha_Q}{24}-\frac{1}{8\pi}\left[\sqrt{\tfrac{11}{6}}\,\frac{3g^3+g'^3}{2}+N_Q\alpha_Q\sqrt{\tfrac{\alpha_Q}{6}+\tfrac{\beta_Q}{24}}\,\right],\\
\left.\frac{\partial^2 V}{\partial s^2}\right|_{0} = c_s T^2,\qquad c_s=\frac{N_Q\beta_Q}{24}\left[1-\frac{3}{\pi}\sqrt{\tfrac{\alpha_Q}{6}+\tfrac{\beta_Q}{24}}\,\right].
\end{gathered}
\label{eq:curv}
\end{equation}
The terms in square brackets come from the daisy resummation, which replaces the $T^2m^2/24$ terms of the light bosons by their resummed counterparts and reduces both curvatures. At BP1 one finds $c_H=0.72$ and $c_s=0.019$, compared with $1.15$ and $0.038$ without the daisy terms, while at BP3 $c_H$ is larger by six percent and $c_s$ smaller by eleven percent. Both remain positive throughout the region of interest, where $\sqrt{\alpha_Q/6+\beta_Q/24}\lesssim0.65$ lies well below $\pi/3$, and $V_0+V_1$ contributes nothing at quadratic order. The origin is therefore a local minimum at every $T>0$, and the barrier separating it from the electroweak vacuum does not disappear as the universe cools. It shrinks with $T$ but never vanishes, which is the origin of the supercooling studied in the rest of the paper. In the conformal limit the bounce action falls only logarithmically with temperature~\cite{Randall:2006py,Konstandin:2011dr}, so the nucleation temperature can lie far below the critical temperature. If no bubbles nucleate before $T\sim\Lambda_{\rm QCD}$, the quark condensate generates a linear term in $H$ and the transition is triggered by QCD dynamics~\cite{Witten:1980ez,Iso:2017uuu,vonHarling:2017yew}. We do not treat that regime and require the transition to complete above $T=1.5$~GeV.

%----------------------------------------------------------------------
\subsection{Classification of thermal histories}
\label{sec:classify}
%----------------------------------------------------------------------

For every parameter point we follow the minima of $V(H,s;T)$ from $T=3$~TeV down to $1.5$~GeV and record two critical temperatures: $T_c^{(s)}$, below which a minimum on the singlet axis $(0,s)$ is lower than the origin, and $T_c$, below which the electroweak minimum $(v(T),v_S(T))$ is. For each candidate first-order step we compute the $O(3)$ bounce action $S_3(T)$ along the two-dimensional tunnelling path with \texttt{CosmoTransitions}~\cite{Wainwright:2011kj}, and define the nucleation temperature $T_n$ by the condition that one bubble nucleates per Hubble volume~\cite{Linde:1981zj},
\begin{equation}
\frac{S_3(T_n)}{T_n} = 4\ln\frac{T_n}{H(T_n)} + \frac{3}{2}\ln\frac{S_3(T_n)}{2\pi T_n},
\qquad
H^2(T)=\frac{1}{3M_{\rm Pl}^2}\left[\frac{\pi^2}{30}g_*T^4+\Delta V_0\right],
\label{eq:Tn}
\end{equation}
where $M_{\rm Pl}=2.435\times10^{18}$~GeV is the reduced Planck mass and $\Delta V_0=V(0,0;0)-V(v,v_S;0)$ is the vacuum energy released by the transition. The vacuum term matters because the Hubble rate is dominated by $\Delta V_0$ once $T$ falls below $(30\Delta V_0/\pi^2g_*)^{1/4}$, which is $36$--$38$~GeV for the benchmarks. The nucleation temperature is a convenient label for the onset of the transition but not for its completion. Percolation and completion, which require tracking the false-vacuum fraction, are the subject of section~\ref{sec:perc}.

\begin{figure}[t]
\centering
\includegraphics[width=0.82\textwidth]{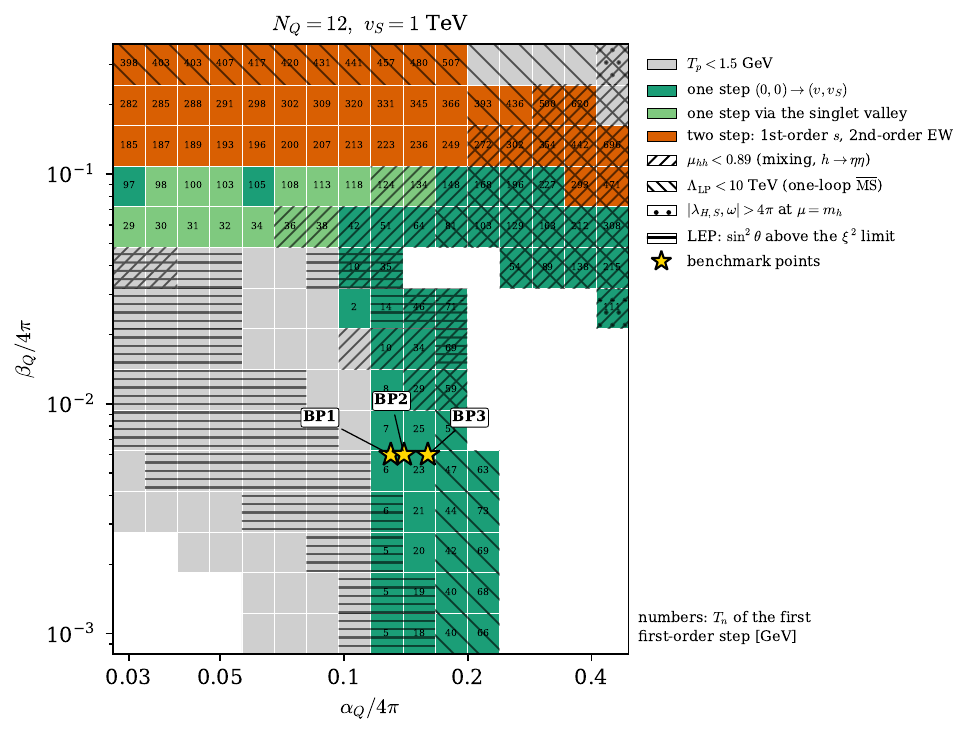}
\caption{Thermal histories in the $(\alpha_Q,\beta_Q)$ plane for $N_Q=12$ and $v_S=1$~TeV. Colours label the sequence of transitions found by tracking the minima of $V(H,s;T)$ from $3$~TeV down to $1.5$~GeV. The number in each cell is the nucleation temperature, in GeV, of the first first-order step. Hatched regions fail the Higgs signal-strength bound, the LEP bound on the dilaton mixing, the Landau-pole condition, or perturbativity of the quartics at $\mu=m_h$ (section~\ref{sec:constraints}). The stars mark BP1--BP3 of table~\ref{tab:bp}.}
\label{fig:history}
\end{figure}

Figure~\ref{fig:history} shows the outcome of this classification at $N_Q=12$ and $v_S=1$~TeV. Five histories occur.
\begin{enumerate}
\item \emph{One step, $(0,0)\to(v,v_S)$.} The singlet axis never develops a minimum below the origin, and the universe tunnels directly into the electroweak vacuum. This is the region of the light-dilaton branch at small $\beta_Q$, and it contains all three benchmarks.
\item \emph{One step via the singlet valley.} The universe first tunnels to a minimum $(0,s)$ on the singlet axis, and the Higgs field condenses within a few percent in temperature of that step. For the purposes of energy release this behaves as a single transition.
\item \emph{Two steps, first-order in $s$ and second-order in $H$.} For $\beta_Q/4\pi\gtrsim0.13$ the singlet acquires its expectation value first, at temperatures of a few hundred GeV, and the electroweak symmetry breaks later through a continuous transition from $(0,s)$.
\item \emph{No completion above $1.5$~GeV.} For $\alpha_Q/4\pi\lesssim0.11$ and $\beta_Q/4\pi\lesssim0.05$ the bounce action stays above the nucleation threshold down to the lowest temperature considered.
\item \emph{No viable vacuum.} The renormalisation conditions~\eqref{eq:rencond} have no solution with $(v,v_S)$ as the global minimum.
\end{enumerate}
The hatched areas are excluded by the constraints of section~\ref{sec:constraints}. The Higgs signal strength removes points where the scalar mixing or the decay $h\to\eta\eta$ is too large, and the LEP bound removes points with large mixing at small $\beta_Q$, including part of the one-step region next to the completion edge. The Landau-pole condition removes the right-hand and upper parts of the plane, where the quartics run fast because $N_Q\alpha_Q^2$ or $N_Q\beta_Q^2$ is large. The one-step region that survives all constraints is a band at $0.12\lesssim\alpha_Q/4\pi\lesssim0.2$ and $\beta_Q/4\pi\lesssim0.03$, whose boundaries in $\alpha_Q$ shift with $\beta_Q$.

The benchmarks sit in the one-step region along $\beta_Q/4\pi=0.006$. Moving along this line from small to large $\alpha_Q$ lowers the amount of supercooling: the critical temperatures are $T_c\simeq 83$, $86$ and $92$~GeV for BP1, BP2 and BP3, while the nucleation temperatures are $T_n\simeq 8.2$, $14.6$ and $30.3$~GeV. At $T_c$ the broken minimum already sits close to its zero-temperature location, with $v(T_c)/T_c\simeq 2.4$--$2.7$ and $v_S(T_c)/T_c\simeq 11$. Increasing $\alpha_Q$ raises both $m_Q$ and the depth of the electroweak vacuum through the $N_Q m_Q^4$ term of eq.~\eqref{eq:mdilaton}, which steepens the potential along the tunnelling direction and lowers $S_3/T$ at a given temperature. Decreasing $\alpha_Q$ has the opposite effect and eventually pushes $T_n$ below $1.5$~GeV, which defines the left edge of the viable one-step region. The singlet portal matters much less. At $\alpha_Q/4\pi=0.152$, for example, $T_n$ rises from $18$ to $29$~GeV as $\beta_Q/4\pi$ grows from $10^{-3}$ to $1.2\times10^{-2}$, while a $20\%$ increase of $\alpha_Q/4\pi$ from $0.127$ to $0.152$ raises it by a factor of four. Section~\ref{sec:NQvS} shows that this edge sits at fixed $N_Q\alpha_Q^2$.

%----------------------------------------------------------------------
\subsection{No first-order electroweak step after singlet breaking}
\label{sec:twostep}
%----------------------------------------------------------------------

Two-step histories have been studied extensively in non-scale-invariant singlet extensions, where an initial singlet transition can prepare a strongly first-order electroweak step. Here they do not. In every point classified as a two-step history, the second step from $(0,s(T))$ to $(v(T),v_S(T))$ is continuous. The reason is visible in the Higgs curvature on the singlet axis,
\begin{equation}
\left.\frac{\partial^2 V}{\partial H^2}\right|_{(0,s)} \simeq c_H T^2 + \frac{\omega}{2}s^2 + \frac{N_Q\alpha_Q\beta_Q}{64\pi^2}\,s^2\left[\ln\frac{\beta_Q s^2}{2\Lambda^2}-1\right],
\label{eq:curvS}
\end{equation}
where the last term is the zero-temperature $Q$ loop and thermal corrections to it are dropped. Once $s$ has condensed, the negative portal $\omega$, which the Gildener-Weinberg flat direction requires, acts as a negative mass term for $H$ and drives this curvature through zero as $T$ decreases. The electroweak step from $(0,s)$ then resembles the Standard Model transition with an effective Higgs mass parameter, where the only cubic term comes from the gauge bosons and is too small to produce a barrier at $m_h=125$~GeV. The Higgs field rolls rather than tunnels. The resulting electroweak transition is second order, or a weak first-order step with $v(T)/T\ll 1$ on our temperature grid, and releases no significant vacuum energy. It therefore produces neither a detectable gravitational-wave signal nor entropy injection.

The phenomenology of the rest of this paper is consequently confined to the one-step region. There the full vacuum energy $\Delta V_0$ is released in a single supercooled transition.

%----------------------------------------------------------------------
\subsection{Tunnelling path and classical roll}
\label{sec:path}
%----------------------------------------------------------------------

\begin{figure}[t]
\centering
\includegraphics[width=\textwidth]{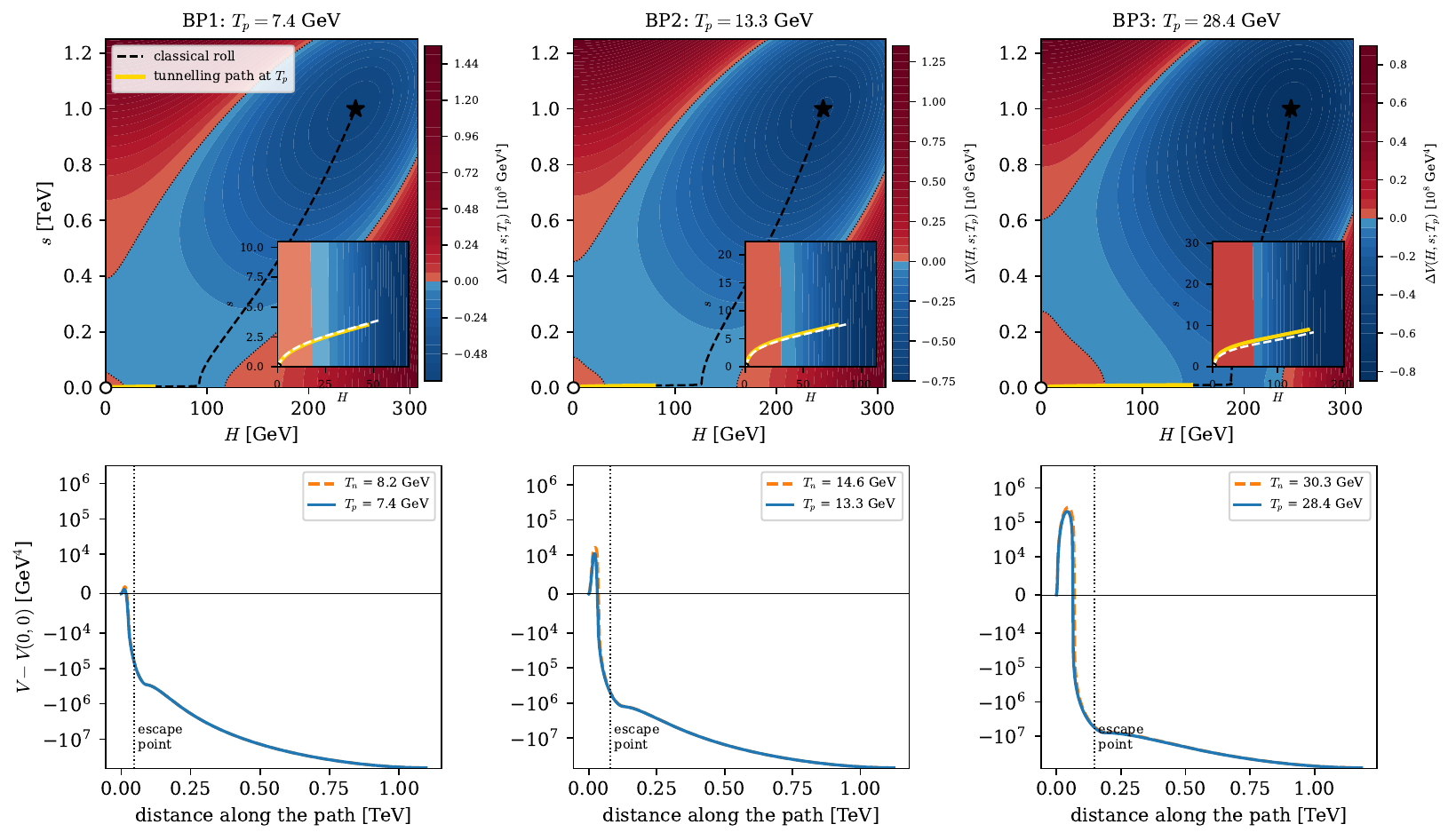}
\caption{Field-space picture of the transition at BP1--BP3. Top: $\Delta V(H,s;T_p)=V(H,s;T_p)-V(0,0;T_p)$, with the tunnelling path at $T_p$ (gold) and the steepest-descent path from its exit point to the minimum (dashed), which indicates the direction of the classical roll but is not a real-time trajectory. The inset magnifies the region near the origin. Bottom: the potential along the tunnelling path followed by the classical roll, at $T_n$ and $T_p$. The marker shows the escape point, where the bounce ends and the classical evolution begins.}
\label{fig:path}
\end{figure}

Figure~\ref{fig:path} shows the field-space structure of the transition at the three benchmarks. The upper panels show $V(H,s;T_p)-V(0,0;T_p)$ at the percolation temperature $T_p$ defined in section~\ref{sec:perc}. The gold curve is the tunnelling path obtained by path deformation, and the dashed curve is the steepest-descent path of the potential from the exit point of the bounce to the minimum. It shows the direction in which the field rolls inside the bubble but is not a real-time trajectory, which would overshoot and oscillate about the valley. The lower panels show the potential along the combined path at $T_n$ and $T_p$.

Two features stand out. First, the bounce is short. At $T_p$ the exit point lies at $(H,s)\simeq(47,4)$, $(79,8)$ and $(148,9)$~GeV for BP1, BP2 and BP3, far from the vacuum at $(246~{\rm GeV},1~{\rm TeV})$. The barrier is thermal and confined to field values of a few times $T$, while the region beyond it is a slope of depth $\Delta V_0^{1/4}\simeq 90$--$96$~GeV. Most of the vacuum energy is therefore released during the classical roll inside the bubble rather than in the tunnelling event itself, which is the field-space counterpart of the large ratio $\alpha$ between vacuum and radiation energy found in section~\ref{sec:energy}.

Second, the path leaves the origin almost exactly along $H$ and turns toward the singlet direction only during the classical roll. The singlet curvature $c_sT^2$ is suppressed by the small $\beta_Q$, but along the singlet axis the potential never drops below its value at the origin, so escaping in that direction gains nothing. Along $H$ the potential falls below the origin a short distance beyond the thermal barrier. The field tunnels there, continues along $H$ to $H\simeq 90$--$190$~GeV, and then turns into the Gildener-Weinberg valley opened by the negative $\omega$ term, which carries it to $(v,v_S)$. This geometry is what makes a single-field description inadequate. A straight path from the origin to the vacuum crosses the barrier where it is widest and overestimates $S_3$ at $T_p$ by a factor of $10$ to $14$ at the benchmarks. The path-deformation solution itself must be started from an initial path that leaves the origin along $H$; started from the straight line it can settle on a critical path whose action is too large by up to a factor of two.

At $T_p$ the bounce actions are $S_3/T\simeq 133$, $134$ and $135$ for BP1, BP2 and BP3. They are nearly equal because the Hubble rate in eq.~\eqref{eq:Tn} is set by the vacuum energy, which varies by only a factor of $1.4$ across the benchmarks, so the nucleation threshold $4\ln(T/H)$ changes only through $\ln T$. The difference between the benchmarks lies in how fast $S_3/T$ decreases with temperature, which fixes how long the universe must supercool to reach this threshold.
 
%======================================================================
\section{Nucleation, percolation and completion}
\label{sec:perc}
%======================================================================

A transition that begins deep in a vacuum-dominated epoch is not guaranteed to end. Bubbles of the true vacuum grow at most at the speed of light, while the false vacuum between them inflates, and if the nucleation rate rises too slowly the false-vacuum volume can grow faster than it is converted~\cite{Guth:1982pn,Turner:1992tz}. The nucleation temperature of section~\ref{sec:classify} only marks the onset of bubble formation. In this section we follow the false-vacuum fraction through the transition, determine the percolation and completion temperatures, and extract the parameters $\alpha$ and $\beta/H$ that control the energy budget and the gravitational-wave signal.

%----------------------------------------------------------------------
\subsection{Nucleation rate and expansion rate}
\label{sec:rate}
%----------------------------------------------------------------------

The nucleation rate per unit volume is~\cite{Linde:1981zj}
\begin{equation}
\Gamma(T) = T^4\left(\frac{S_3(T)}{2\pi T}\right)^{3/2}\exp\!\left[-\frac{S_3(T)}{T}\right],
\label{eq:Gamma}
\end{equation}
where the prefactor approximates the fluctuation determinant. A change of the prefactor by an order of magnitude shifts the nucleation threshold by $\Delta(S_3/T)\simeq2.3$ and $T_p$ by $2.3/(\beta/H)$, about one to two percent at the benchmarks. The expansion rate $H(T)$ is given in eq.~\eqref{eq:Tn}, with $g_*=106.75+1+N_Q$ in the symmetric phase where the singlet and the $Q_i$ are massless. We evaluate $S_3$ along the two-dimensional tunnelling path at $26$ temperatures between $0.97\,T_c$ and $1.5$~GeV and at $11$ further temperatures between $0.8\,T_p$ and $1.3\,T_n$. The path-deformation solver scatters by a few percent from one temperature to the next, so we represent $S_3/T$ by a cubic smoothing spline in $(\ln T,\ln S_3/T)$ that allows a scatter of $\sigma=0.04$ in $\ln(S_3/T)$, slightly above the measured rms of $0.026$ (appendix~\ref{app:numerics}). The quoted uncertainties come from a residual bootstrap of this fit.

\begin{figure}[t]
\centering
\includegraphics[width=\textwidth]{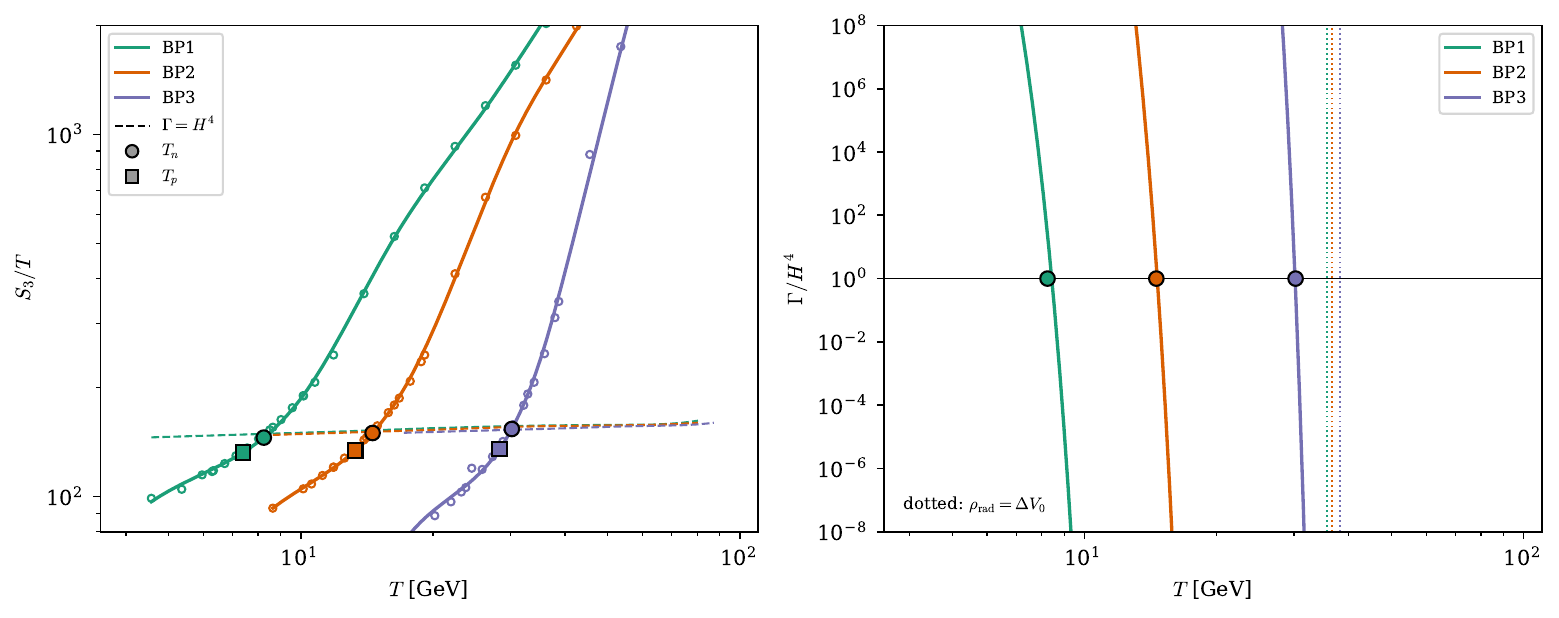}
\caption{Left: $S_3/T$ along the two-dimensional tunnelling path for BP1--BP3. Open circles are individual bounce solutions, solid curves the smoothing-spline fit, and dashed curves the value of $S_3/T$ at which $\Gamma=H^4$, with the vacuum energy included in $H$. Circles mark $T_n$ and squares $T_p$. Right: $\Gamma/H^4$ against temperature. Dotted vertical lines mark $\rho_{\rm rad}=\Delta V_0$.}
\label{fig:rate}
\end{figure}

Figure~\ref{fig:rate} shows the result for the benchmarks. The action falls monotonically as the universe cools, steeply just below $T_c$ and much more slowly at low temperature, where the logarithmic slope at $T_p$ is only $d\ln(S_3/T)/d\ln T\simeq0.9$, $1.2$ and $1.7$ for BP1, BP2 and BP3. The nucleation condition $\Gamma=H^4$ corresponds to $S_3/T\simeq150$ at all three points. This value is set by $\ln(T/H)\simeq36$--$37$, and it barely changes between the benchmarks because the Hubble rate is fixed by the vacuum energy rather than by the temperature. The dotted lines in the right panel mark the temperature $T_{\rm eq}=(30\Delta V_0/\pi^2g_*)^{1/4}\simeq36$--$38$~GeV below which the vacuum energy dominates. BP1 and BP2 nucleate well below it, after about $1.6$ and $1.0$ $e$-folds of vacuum-dominated expansion, while BP3 nucleates close to the boundary.

%----------------------------------------------------------------------
\subsection{Percolation and completion}
\label{sec:percolation}
%----------------------------------------------------------------------

The probability that a point remains in the false vacuum at temperature $T$ is $P_f(T)=e^{-I(T)}$, with~\cite{Guth:1981uk,Turner:1992tz,Ellis:2018mja}
\begin{equation}
I(T) = \frac{4\pi}{3}\int_T^{T_c}\frac{dT'}{T'^4}\,\frac{\Gamma(T')}{H(T')}\left[v_w\int_T^{T'}\frac{dT''}{H(T'')}\right]^3 .
\label{eq:IT}
\end{equation}
Here we have used $a\propto1/T$, which holds in the symmetric phase until the transition because no entropy is produced before it and $g_{*s}$ is constant there, and we set the wall velocity to $v_w=1$. Section~\ref{sec:energy} shows that the walls are ultrarelativistic for $\alpha_Q/4\pi\lesssim0.14$ on the benchmark line, which includes BP1 and BP2. At BP3 the leading-order friction estimate does not apply, and an estimate based on the reflection of the $Q_i$ from the wall gives a Lorentz factor of order five, so that $v_w$ is still close to one. In any case $I(T)$ depends on $v_w$ only through $v_w^3$ and $T_p$ shifts by less than a percent for $v_w$ between $0.5$ and $1$ because $\Gamma$ grows so steeply. We define the percolation temperature $T_p$ by $I(T_p)=0.34$, the threshold for percolation of overlapping spheres of equal size in three dimensions~\cite{Rintoul:1997}, and the completion temperature $T_e$ by $I(T_e)=4.6$, that is $P_f(T_e)=0.01$. At $T_p$ the bubbles form a connected network while $71\%$ of space is still in the false vacuum; $T_e$ is an operational definition of the end of the transition, chosen as the temperature at which only one percent remains. Percolation alone is not sufficient in an inflating background. We also require that the physical volume of the false vacuum decreases at $T_p$~\cite{Turner:1992tz,Ellis:2018mja},
\begin{equation}
\frac{1}{\mathcal{V}_f}\frac{d\mathcal{V}_f}{dt} = H(T)\left[3+T\frac{dI}{dT}\right]_{T_p}<0,
\qquad \mathcal{V}_f\propto a^3P_f .
\label{eq:shrink}
\end{equation}

\begin{figure}[t]
\centering
\includegraphics[width=\textwidth]{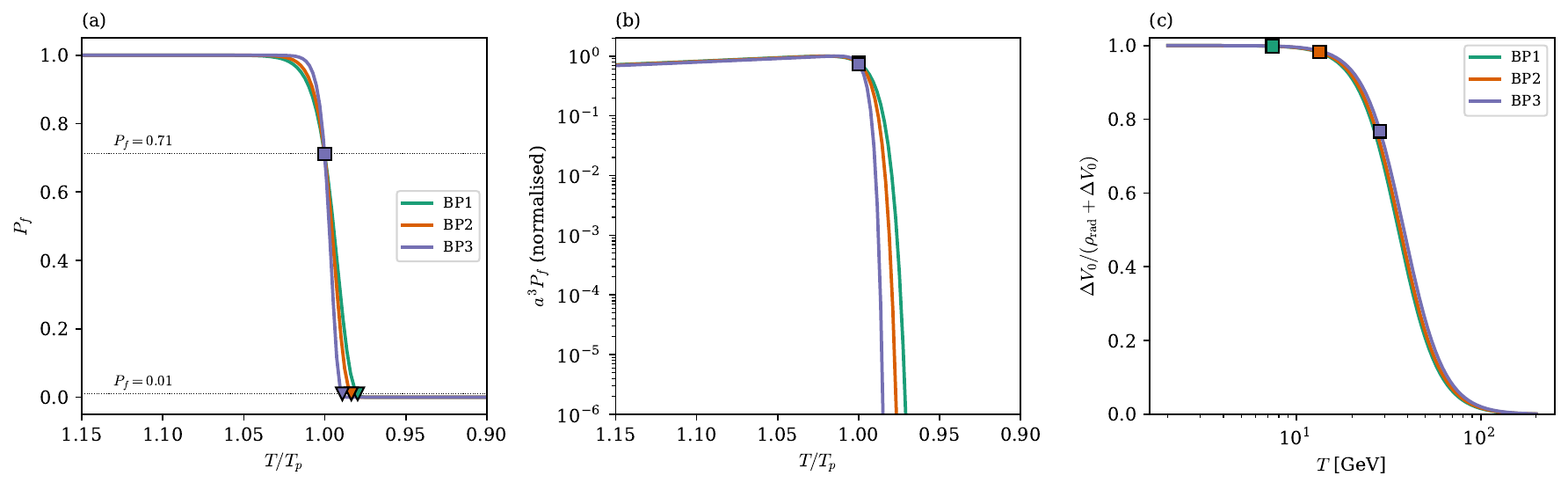}
\caption{Percolation and completion for BP1--BP3. (a) False-vacuum fraction $P_f=e^{-I}$ against $T/T_p$. Squares mark $T_p$ ($P_f=0.71$) and triangles $T_e$ ($P_f=0.01$). (b) Physical false-vacuum volume $a^3P_f$, normalised to its maximum. (c) Fraction of the total energy density carried by the vacuum energy, $\Delta V_0/(\rho_{\rm rad}+\Delta V_0)$, with $T_p$ marked.}
\label{fig:perc}
\end{figure}

Figure~\ref{fig:perc} shows these quantities for the benchmarks. The false-vacuum fraction stays close to one until a few percent above $T_p$ and then collapses. Completion follows percolation within two percent in temperature, $T_e/T_p=0.980$, $0.984$ and $0.989$ for BP1, BP2 and BP3. The bracket in eq.~\eqref{eq:shrink} equals $-54$, $-79$ and $-184$, so the false-vacuum volume shrinks rapidly once bubbles percolate (figure~\ref{fig:perc}b). Since $|T\,dI/dT|$ only grows below $T_p$, the volume keeps decreasing until completion, and there is no indication of regions that remain trapped in the inflating phase. The right panel shows that the vacuum energy accounts for $99.8\%$, $98\%$ and $77\%$ of the total energy density at $T_p$. For BP1 and BP2 the transition therefore takes place in what is effectively a short period of thermal inflation, and the plasma that existed before it has been diluted by the expansion.

The speed with which the transition ends follows from the shape of $S_3/T$. Near $T_p$ the rate grows as $\Gamma\propto e^{\beta t}$ with
\begin{equation}
\frac{\beta}{H} = T\frac{d}{dT}\!\left(\frac{S_3}{T}\right)\bigg|_{T_p} = \frac{S_3(T_p)}{T_p}\,\frac{d\ln(S_3/T)}{d\ln T}\bigg|_{T_p}.
\label{eq:betaH}
\end{equation}
Strong supercooling is often associated with small values of $\beta/H$, but that association holds near a minimum of $S_3/T$, where the logarithmic slope vanishes. Here $S_3/T$ is still decreasing at $T_p$ and its minimum, if any, lies below the temperature range we consider. The slope is of order one, and multiplied by $S_3/T\simeq134$ it gives $\beta/H\simeq130$--$240$ at the benchmarks. A transition with $\beta/H\sim10^2$ lasts about $10^{-2}$ Hubble times, consistent with the narrow interval between $T_p$ and $T_e$.

%----------------------------------------------------------------------
\subsection{Transition parameters}
\label{sec:params}
%----------------------------------------------------------------------

The strength of the transition is measured by the ratio of the released energy to the radiation energy density of the symmetric phase. We use the trace-based definition~\cite{Giese:2020rtr}
\begin{equation}
\alpha = \frac{1}{\rho_{\rm rad}(T_p)}\left[\Delta V(T_p)-\frac{T_p}{4}\frac{\partial\Delta V}{\partial T}\bigg|_{T_p}\right],
\qquad
\rho_{\rm rad}(T)=\frac{\pi^2}{30}g_*T^4 ,
\label{eq:alpha}
\end{equation}
where $\Delta V(T)=V(0,0;T)-V(v(T),v_S(T);T)$. Because the transition proceeds at $T_p\ll T_c$, $\Delta V(T_p)$ is within one percent of the zero-temperature value $\Delta V_0$, and the derivative term changes $\alpha$ by about one percent at BP3 and much less at BP1 and BP2. The strength is then essentially $\alpha\simeq\Delta V_0/\rho_{\rm rad}(T_p)\propto T_p^{-4}$, which explains why it spans two orders of magnitude across the benchmarks while $\Delta V_0$ changes by only $30\%$.

Table~\ref{tab:transition} collects the transition parameters of the benchmarks. The intervals quoted there are the $16$--$84\%$ ranges of the residual bootstrap at fixed smoothing and reflect the numerical scatter of the bounce solver, at the level of one percent in temperature and a few percent in $\alpha$ and $\beta/H$. Two further numerical systematics are not captured by the bootstrap (appendix~\ref{app:numerics}). The choice of the smoothing parameter moves $\beta/H$ by up to twenty percent, and the placement of the temperature grid near $T_p$ moves $T_p$ by up to two percent and $\alpha$ by up to seven percent. The theoretical uncertainty is larger. Varying the renormalisation scale of the thermal potential between $\pi T_p$ and $4\pi T_p$ moves $T_p$ by up to seven percent and $\alpha$ by up to thirty percent, and the choice of resummation prescription moves $T_p$ by about thirteen percent (appendix~\ref{app:scheme}).

\begin{table}[t]
\centering
\renewcommand{\arraystretch}{1.25}
\begin{tabular}{lccc}
\hline\hline
 & BP1 & BP2 & BP3 \\ \hline
$T_c$ [GeV] & 83.0 & 86.0 & 92.1 \\
$T_n$ [GeV] & $8.24^{+0.09}_{-0.08}$ & $14.52^{+0.12}_{-0.12}$ & $30.23^{+0.17}_{-0.18}$ \\
$T_p$ [GeV] & $7.39^{+0.09}_{-0.08}$ & $13.31^{+0.12}_{-0.15}$ & $28.41^{+0.19}_{-0.19}$ \\
$T_e$ [GeV] & $7.24^{+0.09}_{-0.08}$ & $13.10^{+0.13}_{-0.15}$ & $28.12^{+0.20}_{-0.20}$ \\
$S_3(T_p)/T_p$ & 133 & 134 & 135 \\
$\Delta V_0/(\rho_{\rm rad}+\Delta V_0)$ at $T_p$ & 0.998 & 0.983 & 0.768 \\ \hline
$\alpha(T_p)$ & $546^{+25}_{-25}$ & $57^{+3}_{-2}$ & $3.30^{+0.09}_{-0.09}$ \\
$\beta/H(T_p)$ & $127^{+6}_{-6}$ & $168^{+8}_{-13}$ & $243^{+7}_{-8}$ \\
$T_{\rm reh}$ [GeV] & 37.8 & 38.9 & 43.4 \\
$D=s_{\rm after}/s_{\rm before}$ & $106^{+4}_{-4}$ & $20^{+1}_{-1}$ & $2.86^{+0.04}_{-0.04}$ \\
\hline\hline
\end{tabular}
\caption{Transition parameters of the benchmark points. Central values are medians of a residual bootstrap of the fit to $\ln(S_3/T)$, and the quoted intervals are the 16th and 84th percentiles. They reflect the numerical scatter of the bounce solver only. The smoothing of the fit adds an uncertainty of up to $20\%$ to $\beta/H$, and the placement of the temperature grid up to $2\%$ to $T_p$ and $7\%$ to $\alpha$ (appendix~\ref{app:numerics}), and the dependence on the resummation scheme and on the renormalisation scale is discussed in appendix~\ref{app:scheme}. The reheating temperature $T_{\rm reh}$ and the entropy injection $D$ are defined in section~\ref{sec:entropy}.}
\label{tab:transition}
\end{table}

%----------------------------------------------------------------------
\subsection{Characteristic temperatures along the benchmark line}
\label{sec:line}
%----------------------------------------------------------------------

\begin{figure}[t]
\centering
\includegraphics[width=0.75\textwidth]{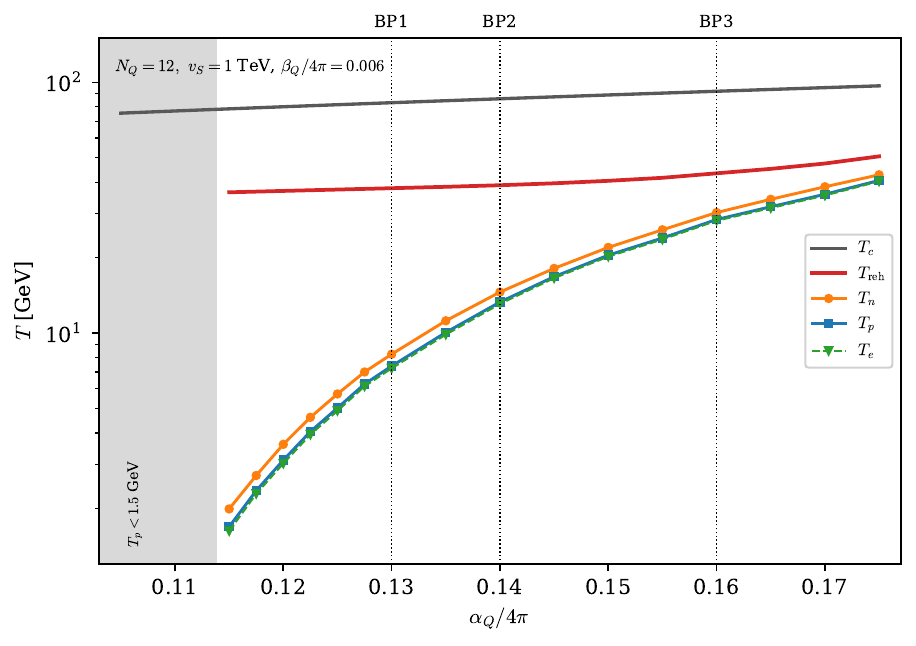}
\caption{Characteristic temperatures along $\beta_Q/4\pi=0.006$ for $N_Q=12$ and $v_S=1$~TeV: critical temperature $T_c$, nucleation temperature $T_n$, percolation temperature $T_p$, completion temperature $T_e$ and reheating temperature $T_{\rm reh}$. The shaded region has $T_p<1.5$~GeV. Dotted lines mark BP1--BP3.}
\label{fig:temps}
\end{figure}

Figure~\ref{fig:temps} follows the characteristic temperatures along the line $\beta_Q/4\pi=0.006$ that contains the benchmarks. The critical temperature changes little, from $75$ to $97$~GeV, because it is fixed by the depth of the electroweak vacuum relative to the thermal free energy, and both grow slowly with $\alpha_Q$. The nucleation and percolation temperatures behave very differently. Over the same range they rise by a factor of twenty, from $T_p\simeq1.7$~GeV at $\alpha_Q/4\pi=0.115$ to $T_p\simeq41$~GeV at $\alpha_Q/4\pi=0.175$. The degree of supercooling $T_p/T_c$ therefore varies from $0.02$ to $0.4$ along the line, and it is this quantity, rather than $T_c$, that controls the transition parameters. The strength falls from $\alpha\simeq2\times10^5$ to $\alpha\simeq0.9$ and $\beta/H$ grows from about $60$ to about $300$.

Completion is never in question along this line. The completion temperature stays within four percent of $T_p$, with $T_e/T_p\geq0.96$, and the bracket of eq.~\eqref{eq:shrink} remains negative, between $-23$ at the lowest $\alpha_Q$ and $-480$ at the highest. The left edge of the viable region, at $\alpha_Q/4\pi\simeq0.114$, is therefore not a failure of percolation in the inflating background. It is the point where $T_p$ reaches $1.5$~GeV, below which the onset of QCD confinement would change the potential and the transition would no longer be described by eq.~\eqref{eq:VT} alone~\cite{Witten:1980ez,Iso:2017uuu,vonHarling:2017yew}. The steep rise of $T_p$ with $\alpha_Q$ near the edge means that the position of this edge depends only weakly on the choice of cutoff.

The reheating temperature, defined in section~\ref{sec:entropy}, is nearly constant at $T_{\rm reh}\simeq36$--$51$~GeV. After strong supercooling the energy that reheats the plasma is almost entirely vacuum energy, so $T_{\rm reh}$ is set by $\Delta V_0^{1/4}$ and hardly depends on how long the universe supercooled. The plasma is reheated above $T_p$ at every point, by a factor of up to twenty near the edge, and the gap between $T_{\rm reh}$ and $T_p$ is the origin of the entropy injection discussed in section~\ref{sec:entropy}.

%======================================================================
\section{Energy budget}
\label{sec:energy}
%======================================================================

The vacuum energy released at $T_p$ ends up in three places: the gradient and kinetic energy of the scalar field in the bubble walls, bulk motion of the plasma, and heat. Which of these dominates decides both the shape of the gravitational-wave spectrum and how quickly the plasma thermalises after the transition. For transitions with $\alpha\gg1$ the answer depends on whether the plasma can stop the walls from accelerating indefinitely, and we address this question here before turning to the spectrum in section~\ref{sec:gw}.

%----------------------------------------------------------------------
\subsection{Strength and duration along the benchmark line}
\label{sec:alphabeta}
%----------------------------------------------------------------------

\begin{figure}[t]
\centering
\includegraphics[width=\textwidth]{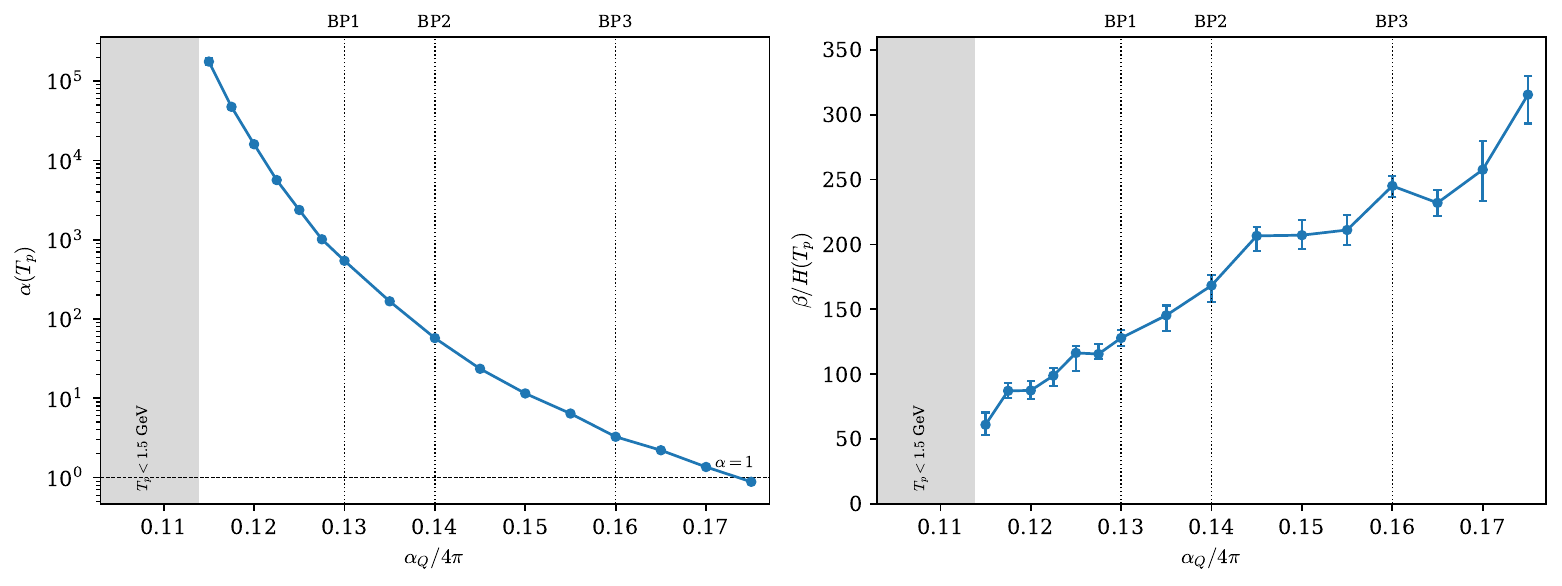}
\caption{Transition strength $\alpha(T_p)$ (left) and inverse duration $\beta/H(T_p)$ (right) along $\beta_Q/4\pi=0.006$, for $N_Q=12$ and $v_S=1$~TeV. Points are bootstrap medians and error bars the 16th to 84th percentile intervals of the fit to $\ln(S_3/T)$. The shaded region has $T_p<1.5$~GeV and dotted lines mark BP1--BP3.}
\label{fig:alphabeta}
\end{figure}

Figure~\ref{fig:alphabeta} shows $\alpha$ and $\beta/H$ along the line $\beta_Q/4\pi=0.006$. The strength falls by more than five orders of magnitude, from $\alpha\simeq1.8\times10^5$ at the completion edge to $\alpha\simeq0.9$ at $\alpha_Q/4\pi=0.175$. As noted in section~\ref{sec:params}, almost all of this variation comes from $\rho_{\rm rad}(T_p)\propto T_p^4$ and not from the released energy, which grows only by a factor of $1.7$ along the line. The inverse duration grows from $\beta/H\simeq60$ to $\simeq320$. Its increase reflects the steepening of $S_3/T$ near $T_p$ at larger $\alpha_Q$, since $S_3(T_p)/T_p$ itself stays between $130$ and $135$ along the whole line (eq.~\eqref{eq:betaH}).

The error bars in figure~\ref{fig:alphabeta} are the bootstrap intervals of the fit and are a few percent for $\alpha$ and about $5\%$ for $\beta/H$. The scatter of $\beta/H$ between neighbouring points is somewhat larger, up to about $10\%$, because $\beta/H$ depends on the derivative of the fitted action and is therefore more sensitive to the residual scatter of individual bounce solutions than $T_p$ or $\alpha$. Together with the dependence on the smoothing of the fit, studied in appendix~\ref{app:numerics}, this leads us to assign a numerical uncertainty of about $20\%$ to $\beta/H$.

%----------------------------------------------------------------------
\subsection{Friction on ultrarelativistic walls}
\label{sec:friction}
%----------------------------------------------------------------------

A wall moving through the plasma with Lorentz factor $\gamma\gg1$ experiences a pressure from particles that gain mass as they cross it. At leading order this pressure saturates at a value independent of $\gamma$~\cite{Bodeker:2009qy},
\begin{equation}
\frac{\Delta P_{\rm LO}}{\rho_{\rm rad}} = \alpha_\infty = \frac{T_p^2}{24\,\rho_{\rm rad}(T_p)}\sum_i c_i\,n_i\,\Delta m_i^2 ,
\label{eq:ainf}
\end{equation}
where $\Delta m_i^2$ is the change of the squared mass of species $i$ across the wall and $c_i=1$ ($1/2$) for bosons (fermions). In the present model all masses vanish in the symmetric phase, so $\sum_i c_in_i\Delta m_i^2 = 6m_W^2+3m_Z^2+6m_t^2+N_Qm_Q^2+m_h^2+m_\eta^2$, dominated by the $Q_i$. We evaluate $\Delta m_i^2$ with the zero-temperature masses; at $T_p$ the broken minimum lies within one percent of $(v,v_S)$, so the difference is negligible. Equation~\eqref{eq:ainf} also assumes that the particles are energetic enough to enter the broken phase, $\gamma T_p\gg m_Q$, a condition we check below. If $\alpha<\alpha_\infty$ the leading-order pressure alone stops the acceleration, and the wall velocity is set by the full non-equilibrium dynamics. If $\alpha>\alpha_\infty$ the wall keeps accelerating until the next-to-leading-order pressure from transition radiation of gauge bosons, which grows with $\gamma$, balances the remaining driving force~\cite{Bodeker:2017cim},
\begin{equation}
\frac{\Delta P_{\rm NLO}}{\rho_{\rm rad}} = \gamma\,\alpha_{\rm eq},\qquad
\alpha_{\rm eq} = \frac{T_p^3}{\rho_{\rm rad}(T_p)}\left[6g^2m_W+3(g^2+g'^2)m_Z\right].
\label{eq:aeq}
\end{equation}
For $\alpha>\alpha_\infty$ the terminal Lorentz factor is then~\cite{Ellis:2019oqb}
\begin{equation}
\gamma_{\rm eq} = \frac{\alpha-\alpha_\infty}{\alpha_{\rm eq}} .
\label{eq:geq}
\end{equation}
Here $\alpha$ stands for the driving pressure $\Delta V(T_p)/\rho_{\rm rad}(T_p)$, which differs from the trace-based $\alpha$ of eq.~\eqref{eq:alpha} by at most $1.2\%$ at the benchmarks.
In the absence of friction a wall would instead reach $\gamma_*=2R_*/(3R_0)$ by the time bubbles collide, where $R_0$ is the radius of the critical bubble, which we take as the radius at which the field falls to half its central value, and $R_*=(8\pi)^{1/3}v_w/\beta$ is the mean bubble separation~\cite{Ellis:2019oqb}.

Table~\ref{tab:budget} lists these quantities at the benchmarks. The ratio $\alpha/\alpha_\infty$ scales as $T_p^{-2}$, because $\alpha_\infty\propto T_p^{-2}$ while $\alpha\propto T_p^{-4}$, so stronger supercooling makes the walls more likely to run away at leading order. At BP1 and BP2 the driving force exceeds the leading-order pressure by factors of $22$ and $7$, and the walls reach $\gamma_{\rm eq}\simeq430$ and $73$. The condition $\gamma T_p\gg m_Q$ is then satisfied, with $\gamma_{\rm eq}T_p/m_Q\simeq11$ and $3.2$, although only marginally at BP2. At BP3 eq.~\eqref{eq:geq} would give $\gamma_{\rm eq}\simeq4.5$, but then $\gamma_{\rm eq}T_p/m_Q\simeq0.4$: a typical $Q$ in the wall frame does not have enough energy to enter the broken phase, and eq.~\eqref{eq:ainf} does not apply. In this regime most $Q_i$ are reflected, and the reflected flux exerts a pressure of order $\gamma^2\rho_Q$, with $\rho_Q$ the energy density of the $Q_i$ in the symmetric phase. Balancing it against the driving pressure gives $\gamma\sim(\alpha\,g_*/N_Q)^{1/2}\approx4$--$6$ at BP3, where the range reflects the order-one factor in the reflection pressure. The wall velocity, $v_w\simeq0.97$--$0.99$, is close to the Jouguet velocity $v_J\simeq0.976$, where the efficiency factor of eq.~\eqref{eq:ksw} would lie between its $v_w\to1$ value of $0.79$ and its value $0.84$ at $v_J$~\cite{Espinosa:2010hh}. The resulting uncertainty in the kinetic fraction at BP3 is below ten percent. In all three cases $\gamma_*\sim10^{12}$, so the walls reach their terminal velocity when the bubbles are smaller than $10^{-9}$ of their final size, and they spend essentially the whole transition at constant velocity.

Along the benchmark line $\gamma_{\rm eq}$ decreases from $3\times10^4$ at the completion edge to $2.4$ at $\alpha_Q/4\pi=0.165$, and the ratio $\gamma_{\rm eq}T_p/m_Q$ falls from $11$ at $\alpha_Q/4\pi=0.13$ to $3.2$ at $0.14$ and $1.2$ at $0.15$. The leading-order estimate of the wall velocity is therefore self-consistent only for $\alpha_Q/4\pi\lesssim0.14$. Between $0.15$ and $0.165$, which includes BP3, the reflection estimate above gives walls with Lorentz factors of a few. For $\alpha_Q/4\pi\gtrsim0.17$ the transitions are weakest ($\alpha\lesssim1.5$), our estimates do not determine $v_w$, and the walls may well reach a terminal velocity below the speed of light; we keep $v_w=1$ there only as an assumption.

The scaling of the next-to-leading-order pressure with $\gamma$ is not settled. A resummation of soft emissions gives $\Delta P\propto\gamma^2$ rather than $\gamma$~\cite{Hoche:2020ysm}, while other analyses retain the linear growth~\cite{Gouttenoire:2021kjv}. A faster growth of the pressure would lower $\gamma_{\rm eq}$ further and strengthen the conclusion of the next subsection.

\begin{table}[t]
\centering
\renewcommand{\arraystretch}{1.2}
\begin{tabular}{lccc}
\hline\hline
 & BP1 & BP2 & BP3 \\ \hline
$\alpha$ & 546 & 57 & 3.31 \\
$\alpha_\infty$ & 25.2 & 8.1 & 1.9 \\
$\alpha_{\rm eq}$ & 1.21 & 0.67 & 0.31 \\
$\gamma_{\rm eq}$ & 431 & 73 & $4.5^{\dagger}$ \\
$\gamma_{\rm eq}T_p/m_Q$ & 11 & 3.2 & 0.4 \\
$1-v_w$ & $2.7\times10^{-6}$ & $9.3\times10^{-5}$ & $(1.4$--$3.2)\times10^{-2\,\dagger}$ \\
\hline
$R_0T_p$ & 9.7 & 10.7 & 11.2 \\
$R_*H$ & 0.0231 & 0.0174 & 0.0121 \\
$\gamma_*=2R_*/3R_0$ & $6.2\times10^{12}$ & $7.2\times10^{12}$ & $8.2\times10^{12}$ \\
\hline
$\kappa_{\rm col}$ & $7.0\times10^{-11}$ & $1.0\times10^{-11}$ & $5.5\times10^{-13}$ \\
$\kappa_{\rm sw}$ & 0.995 & 0.977 & 0.789 \\
$K=\kappa_{\rm sw}\alpha/(1+\alpha)$ & 0.993 & 0.960 & 0.606 \\
\hline\hline
\end{tabular}
\caption{Energy budget of the benchmark points at $T_p$. $\alpha_\infty$ and $\alpha_{\rm eq}$ are the leading-order and next-to-leading-order friction parameters of eqs.~\eqref{eq:ainf} and~\eqref{eq:aeq}, $\gamma_{\rm eq}$ the terminal Lorentz factor of the walls from eq.~\eqref{eq:geq}, $R_0$ the radius of the critical bubble, $R_*$ the mean bubble separation and $\gamma_*$ the Lorentz factor the walls would reach at collision in the absence of friction. $\kappa_{\rm col}$ and $\kappa_{\rm sw}$ are the fractions of the released energy carried by the scalar field and by bulk fluid motion, and $K$ is the fraction of the total energy density in bulk fluid motion. $^\dagger$At BP3 $\gamma_{\rm eq}T_p<m_Q$ and the leading-order friction estimate does not apply; the entry for $1-v_w$ uses the reflection estimate $\gamma\approx4$--$6$ of section~\ref{sec:friction}.}
\label{tab:budget}
\end{table}

%----------------------------------------------------------------------
\subsection{Efficiency factors}
\label{sec:kappa}
%----------------------------------------------------------------------

Once the walls have reached $\gamma_{\rm eq}\ll\gamma_*$, the energy stored in them stops growing with the bubble radius, and the released vacuum energy goes into the plasma. The fraction carried by the scalar field at collision is~\cite{Ellis:2019oqb}
\begin{equation}
\kappa_{\rm col} = \frac{\gamma_{\rm eq}}{\gamma_*}\left[1-\frac{\alpha_\infty}{\alpha}\left(\frac{\gamma_{\rm eq}}{\gamma_*}\right)^2\right]
\qquad (\gamma_*>\gamma_{\rm eq}),
\label{eq:kcol}
\end{equation}
which is below $10^{-10}$ at the benchmarks. Since $\kappa_{\rm col}$ is so small, the convention used for $R_0$ has no effect on the results. Bubble collisions therefore make no relevant contribution to the gravitational-wave signal, in contrast to scenarios in which the walls run away~\cite{Ellis:2019oqb}. The fraction of the released energy converted into bulk fluid motion is~\cite{Espinosa:2010hh}
\begin{equation}
\kappa_{\rm sw} = \left(1-\kappa_{\rm col}\right)\frac{\alpha}{0.73+0.083\sqrt{\alpha}+\alpha},
\label{eq:ksw}
\end{equation}
which is the fit for $v_w\to1$. It gives $\kappa_{\rm sw}\simeq0.995$, $0.977$ and $0.79$ at BP1, BP2 and BP3, and the fraction of the total energy density in bulk motion, $K=\kappa_{\rm sw}\alpha/(1+\alpha)$, is $0.99$, $0.96$ and $0.61$.

These numbers say that for the strongest transitions almost the entire energy density of the universe is converted into the kinetic energy of thin, ultrarelativistic shells of plasma that move ahead of the walls. Equation~\eqref{eq:ksw} was obtained for a stationary self-similar profile and is an extrapolation at $\alpha\gg1$. The qualitative picture, in which the energy is in the fluid rather than in the scalar field, does not depend on that extrapolation, but how efficiently such shells source a long-lived acoustic signal is an open question. Lattice simulations of the acoustic signal reach $\alpha$ of order one~\cite{Cutting:2019zws}, while for $\alpha\gg1$ the shells have been modelled as thin relativistic sources whose spectrum resembles that of bubble collisions~\cite{Lewicki:2022pdb}. We return to this point when estimating the spectrum in section~\ref{sec:gw}. The bulk kinetic energy is not yet heat. It is dissipated by shocks and turbulence on the time scale $\tau_{\rm sw}=R_*/\bar U_f$ of section~\ref{sec:gw}, a few times $1/\beta$ and thus about one percent of a Hubble time, after which the plasma is the reheated radiation bath of section~\ref{sec:entropy}.
 
%======================================================================
\section{Entropy injection and dilution of relics}
\label{sec:entropy}
%======================================================================

A supercooled transition behaves in many respects like a short period of thermal inflation~\cite{Lyth:1995ka}. While the universe is trapped in the symmetric phase, the radiation bath redshifts away and the vacuum energy stays constant. When the transition completes, that vacuum energy is converted into radiation at a temperature set by $\Delta V_0^{1/4}$, independently of how cold the plasma had become. Any species that was already decoupled before the transition does not share in this reheating, and its abundance relative to the entropy density is reduced. This section quantifies the effect and its consequences for relics produced earlier.

%----------------------------------------------------------------------
\subsection{Reheating temperature and entropy injection}
\label{sec:reheat}
%----------------------------------------------------------------------

Because the transition completes within about $10^{-2}$ Hubble times (section~\ref{sec:percolation}), we treat reheating as instantaneous and impose energy conservation at $T_p$,
\begin{equation}
\frac{\pi^2}{30}g_*^{\rm f}T_p^4 + \Delta\rho(T_p) = \frac{\pi^2}{30}g_*^{\rm b}(T_{\rm reh})\,T_{\rm reh}^4,
\qquad
\Delta\rho = \Delta V - T\frac{\partial\Delta V}{\partial T},
\label{eq:reheat}
\end{equation}
where $g_*^{\rm f}=106.75+1+N_Q$ counts the massless degrees of freedom of the symmetric phase and $g_*^{\rm b}(T)$ is evaluated from the broken-phase spectrum, in which the $Q_i$ ($m_Q\simeq300$~GeV), the top quark and the electroweak bosons are Boltzmann suppressed at $T_{\rm reh}\sim40$~GeV; at the benchmarks $g_{*s}^{\rm b}(T_{\rm reh})\simeq94.6$, $94.9$ and $96.2$. The released energy is assumed to thermalise into the Standard Model bath. This is justified because the energy is carried by bulk motion of that same plasma, which is dissipated within a few times $1/\beta$, about one percent of a Hubble time (section~\ref{sec:kappa}). The ratio of the entropy densities after and before the transition is then
\begin{equation}
D \equiv \frac{s_{\rm after}}{s_{\rm before}} = \frac{g_{*s}^{\rm b}(T_{\rm reh})\,T_{\rm reh}^3}{g_{*s}^{\rm f}\,T_p^3}
\simeq \left(\frac{g_*^{\rm b}}{g_*^{\rm f}}\right)^{1/4}\left(1+\alpha\right)^{3/4},
\label{eq:D}
\end{equation}
where the approximate form uses $\Delta\rho\simeq\alpha\,\rho_{\rm rad}(T_p)$ and $g_{*s}\simeq g_*$. It reproduces the full result to within $3\%$ at the benchmarks.

The values are listed in table~\ref{tab:transition}. The reheating temperature is close to $T_{\rm eq}=(30\Delta V_0/\pi^2g_*)^{1/4}$ for all supercooled points, $T_{\rm reh}\simeq38$, $39$ and $43$~GeV at BP1, BP2 and BP3, while the entropy injection ranges from $D\simeq106$ at BP1 to $D\simeq2.9$ at BP3. Along the benchmark line, $D$ grows from $1.6$ at $\alpha_Q/4\pi=0.175$ to $8\times10^3$ at the completion edge (figure~\ref{fig:dilution}a).

Two consistency conditions are satisfied throughout. First, $T_{\rm reh}<T_c$, so the plasma is reheated into the broken phase and the transition is not undone. At $T_{\rm reh}$ the Higgs expectation value satisfies $v(T_{\rm reh})/T_{\rm reh}\simeq5.7$--$6.5$ at the benchmarks, far above the threshold $v/T\gtrsim1$ for sphaleron decoupling. Second, $T_{\rm reh}$ is many orders of magnitude above the MeV scale, so the standard thermal history below $T_{\rm reh}$, including big-bang nucleosynthesis, is unaffected.

%----------------------------------------------------------------------
\subsection{Entropy injection across the parameter space}
\label{sec:Dplane}
%----------------------------------------------------------------------

\begin{figure}[t]
\centering
\includegraphics[width=0.78\textwidth]{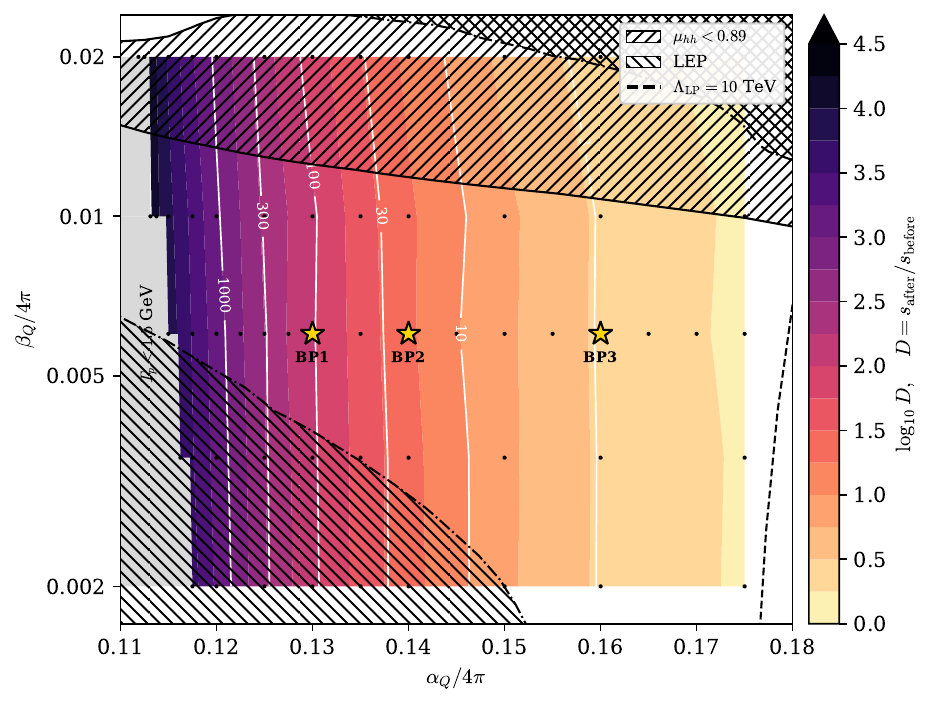}
\caption{Entropy injection $D=s_{\rm after}/s_{\rm before}$ in the $(\alpha_Q,\beta_Q)$ plane for $N_Q=12$ and $v_S=1$~TeV. Colours and white contours give $D$, interpolated along the five computed lines of fixed $\beta_Q$ (dots). The grey region has $T_p<1.5$~GeV, the region hatched to the right has $\mu_{hh}<0.89$, the region hatched to the left and bounded by the dash-dotted line is excluded by LEP, and the dashed line marks $\Lambda_{\rm LP}=10$~TeV. Stars mark BP1--BP3.}
\label{fig:Dplane}
\end{figure}

Figure~\ref{fig:Dplane} shows $D$ over the part of the $(\alpha_Q,\beta_Q)$ plane where the transition is a single supercooled step. The contours are almost vertical, so the entropy injection is controlled by $\alpha_Q$. At $\alpha_Q/4\pi=0.13$ it varies only between $D\simeq105$ and $113$ for $\beta_Q/4\pi$ between $0.002$ and $0.01$, although the LEP bound excludes the part of this range below $\beta_Q/4\pi\simeq0.004$. The singlet portal acts mainly through the position of the completion edge, which moves from $\alpha_Q/4\pi\simeq0.1175$ at $\beta_Q/4\pi=0.002$ to $\alpha_Q/4\pi\simeq0.113$ at $\beta_Q/4\pi=0.01$, and through the vacuum energy. A larger $\beta_Q$ increases $m_Q$ at fixed $\alpha_Q$ and hence $\Delta V_0$, which raises the reheating temperature at the edge from $30$ to $42$~GeV over the same range of $\beta_Q$.

The entropy injection is largest along the completion edge, where $T_p$ approaches $1.5$~GeV. There it reaches $D\simeq4\times10^3$ at $\beta_Q/4\pi=0.002$ and $D\simeq1.5\times10^4$ at $\beta_Q/4\pi=0.01$. The first of these lies in the region excluded by LEP, which removes the edge for $\beta_Q/4\pi\lesssim0.005$, while the second is allowed with $\sin^2\theta$ at half the LEP limit. Because $D\propto\Delta V_0^{3/4}/T_p^3$ for $\alpha\gg1$, the edge value depends on the cutoff we impose on $T_p$ and would grow further if transitions triggered by QCD effects were included. Above $\beta_Q/4\pi\simeq0.01$--$0.015$, depending on $\alpha_Q$, the Higgs signal strength excludes the plane, and on the right the Landau-pole condition sets in at $\alpha_Q/4\pi\simeq0.177$, where $D$ has already fallen to about $1.5$. The viable region therefore supports entropy injection over four orders of magnitude, $1.5\lesssim D\lesssim10^4$, with $D\gtrsim100$ for $\alpha_Q/4\pi\lesssim0.13$.

%----------------------------------------------------------------------
\subsection{Consequences for relics}
\label{sec:relics}
%----------------------------------------------------------------------

\begin{figure}[t]
\centering
\includegraphics[width=\textwidth]{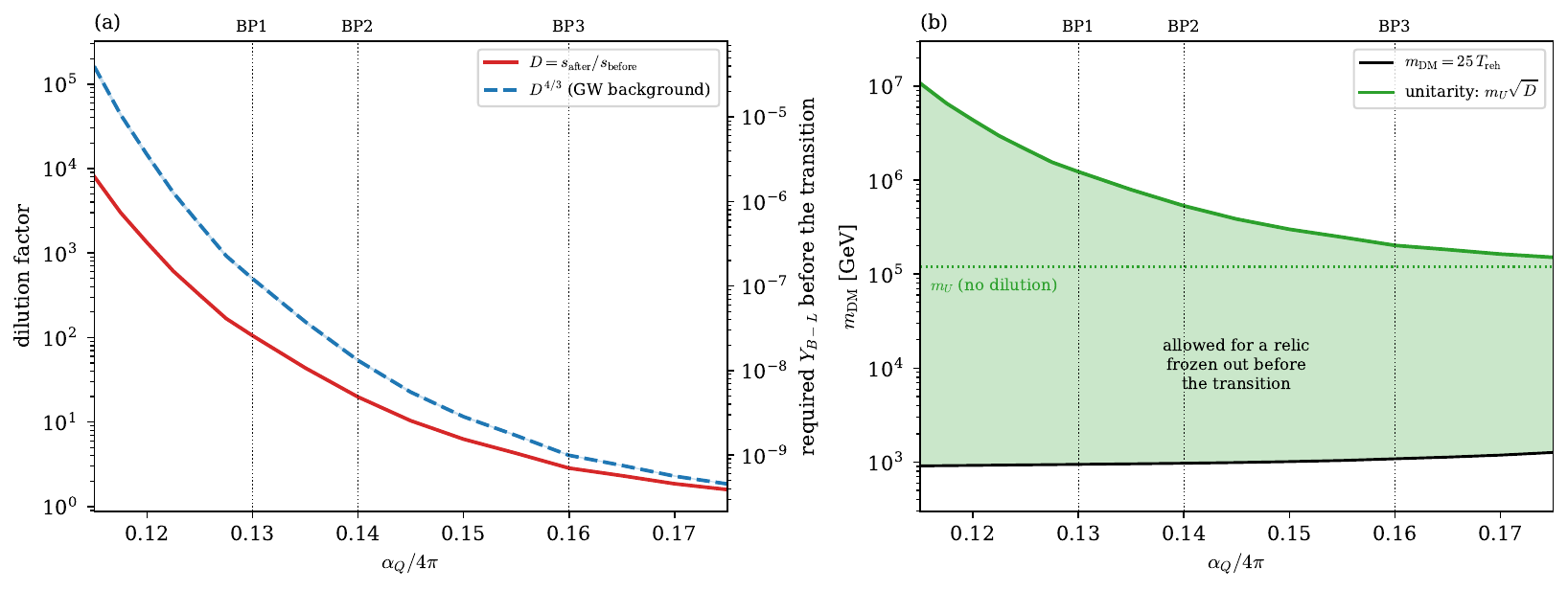}
\caption{Dilution of pre-existing relics along $\beta_Q/4\pi=0.006$. (a) Entropy injection $D$ (solid) and the dilution $D^{4/3}$ of a pre-existing gravitational-wave background (dashed). The right axis gives the $B-L$ asymmetry required before the transition to reproduce the observed baryon asymmetry, eq.~\eqref{eq:YBL}. (b) Mass window for a thermal relic that freezes out before the transition: the lower edge, $m_{\rm DM}=25\,T_{\rm reh}$, avoids re-equilibration after reheating, and the upper edge is the unitarity bound $m_U\sqrt{D}$. Bands show the bootstrap intervals and dotted lines mark BP1--BP3.}
\label{fig:dilution}
\end{figure}

The abundance $Y_X=n_X/s$ of any species $X$ that is out of equilibrium during the transition and is not produced by the reheated plasma is reduced by $D$. Energy densities that redshift like radiation, such as a pre-existing gravitational-wave background, are reduced relative to the radiation bath by $D^{4/3}$, up to the change in the number of relativistic degrees of freedom. The $Q_i$ themselves are not relics of this kind: they return to equilibrium after reheating and decay before nucleosynthesis through the couplings of eq.~\eqref{eq:kappa}. Figure~\ref{fig:dilution} illustrates three consequences along the benchmark line.

\paragraph{Baryon asymmetry.}
The symmetric phase persists down to $T_p\ll T_c$, and in this phase electroweak sphalerons remain in equilibrium. Their rate per unit volume is $\Gamma_{\rm sph}\simeq18\,\alpha_w^5T^4$~\cite{DOnofrio:2014kta}, so $\Gamma_{\rm sph}/T^3\sim10^{-5}$~GeV at $T\sim10$~GeV, compared with $H\sim10^{-15}$~GeV set by the vacuum energy. Any baryon asymmetry generated before the transition is therefore reprocessed into $B=\tfrac{28}{79}(B-L)$~\cite{Khlebnikov:1988sr,Harvey:1990qw}, and only $B-L$ survives to $T_p$. After the transition the plasma is in the broken phase with $v/T\gg1$, sphalerons are frozen out, and the baryon number is diluted but not washed out. Reproducing the observed value $Y_B\simeq8.7\times10^{-11}$~\cite{Planck:2018vyg} requires an initial asymmetry
\begin{equation}
Y_{B-L}^{\rm ini} = \frac{79}{28}\,D\,Y_B^{\rm obs},
\label{eq:YBL}
\end{equation}
shown on the right axis of figure~\ref{fig:dilution}a. At BP1 this is $Y_{B-L}^{\rm ini}\simeq2.6\times10^{-8}$, about $300$ times the observed asymmetry, and near the completion edge it reaches $2\times10^{-6}$. Asymmetries this large are not generic, so the strongest transitions constrain any baryogenesis mechanism that operates before the electroweak scale. Conversely, a mechanism that overproduces the asymmetry is brought into agreement with observation.

\paragraph{Thermal relics.}
A stable particle that freezes out before the transition and is not regenerated afterwards has its relic density reduced by $D$, so the annihilation cross section required to reproduce the observed dark-matter abundance is smaller by the same factor~\cite{Hambye:2018qjv}. Two conditions bound the mass of such a relic. The reheated plasma must not bring it back into equilibrium. We estimate this with the criterion $m_{\rm DM}\gtrsim25\,T_{\rm reh}\simeq1$~TeV appropriate for a weakly interacting particle; the precise threshold depends on how strongly the relic couples to the plasma. In addition, perturbative unitarity caps the $s$-wave annihilation cross section and in standard cosmology implies $m_{\rm DM}\lesssim m_U\simeq1.2\times10^5$~GeV~\cite{Griest:1989wd}. Since $\Omega_{\rm DM}\propto m_{\rm DM}^2$ at the unitarity limit, dilution relaxes this bound to $m_{\rm DM}\lesssim m_U\sqrt{D}$. Figure~\ref{fig:dilution}b shows the resulting window. At BP1 thermal relics up to about $1.2\times10^6$~GeV become viable, and near the completion edge the bound rises above $10^7$~GeV. The value of $m_U$ depends on the spin of the relic and on the annihilation channel, and we use it only to indicate the scale.

\paragraph{Earlier gravitational-wave backgrounds.}
A stochastic background produced before the transition, for example by cosmic strings or by an earlier phase transition, has its energy density relative to radiation reduced by $D^{4/3}$ (figure~\ref{fig:dilution}a). For BP1 this is a factor of about $500$. The signal from the transition itself is produced at $T_p$ and redshifts with the reheated plasma afterwards, so it is not affected by its own entropy injection.

%======================================================================
\section{Gravitational waves}
\label{sec:gw}
%======================================================================

Section~\ref{sec:energy} established that the released energy goes almost entirely into bulk motion of the plasma, so the gravitational-wave signal is sourced by the fluid rather than by the scalar field. In this section we estimate the resulting spectrum, show that its peak follows an approximate scaling relation set by the vacuum energy, and assess its detectability with LISA and DECIGO.

%----------------------------------------------------------------------
\subsection{Acoustic spectrum}
\label{sec:swspec}
%----------------------------------------------------------------------

Sound waves were identified as the dominant and longest-lasting source in simulations of the coupled scalar-fluid system~\cite{Hindmarsh:2013xza,Hindmarsh:2015qta}. We use the sound-wave template of refs.~\cite{Hindmarsh:2017gnf,Caprini:2019egz},
\begin{align}
h^2\Omega_{\rm sw}(f) &= 4.13\times10^{-7}\,(R_*H_*)\,\Upsilon\left(\frac{\kappa_{\rm sw}\alpha}{1+\alpha}\right)^2\left(\frac{100}{g_*}\right)^{1/3}S\!\left(\frac{f}{f_p}\right),
\qquad S(x)=x^3\left(\frac{7}{4+3x^2}\right)^{7/2},
\label{eq:Osw}\\
f_p &= 2.6\times10^{-5}\,{\rm Hz}\,\frac{1}{R_*H_*}\left(\frac{T_*}{100~{\rm GeV}}\right)\left(\frac{g_*}{100}\right)^{1/6},
\label{eq:fp}
\end{align}
with $R_*=(8\pi)^{1/3}v_w/\beta$. The factor
\begin{equation}
\Upsilon = 1-\frac{1}{\sqrt{1+2\tau_{\rm sw}H_*}},\qquad
\tau_{\rm sw}=\frac{R_*}{\bar U_f},\qquad
\bar U_f^2=\frac{3}{4}\frac{\kappa_{\rm sw}\alpha}{1+\alpha},
\label{eq:Ups}
\end{equation}
accounts for the finite lifetime of the acoustic source, which ends when nonlinearities develop after a time $\tau_{\rm sw}$~\cite{Ellis:2020awk,Guo:2020grp}. These expressions were calibrated on simulations with $\alpha\lesssim0.1$ and subrelativistic walls. For $\alpha\gg1$ and ultrarelativistic fluid shells they are an extrapolation (section~\ref{sec:kappa}), so the amplitudes and signal-to-noise ratios below are estimates that hold if the acoustic description carries over to this regime.

Two choices specific to supercooled transitions enter these expressions. The spectrum is produced after the plasma has been reheated, so the temperature and number of degrees of freedom that set the redshift are $T_*=T_{\rm reh}$ and $g_*=g_*^{\rm b}(T_{\rm reh})$ rather than their values at $T_p$. The Hubble rate $H_*$ is the same before and after reheating, because the energy density does not change in the conversion of vacuum energy into radiation, and it is dominated by $\Delta V_0$. With these choices the spectrum depends on the transition through only three quantities: $\beta/H$, the kinetic fraction $K=\kappa_{\rm sw}\alpha/(1+\alpha)$, and $T_{\rm reh}$.

%----------------------------------------------------------------------
\subsection{Benchmark spectra}
\label{sec:bpspec}
%----------------------------------------------------------------------

\begin{figure}[t]
\centering
\includegraphics[width=0.85\textwidth]{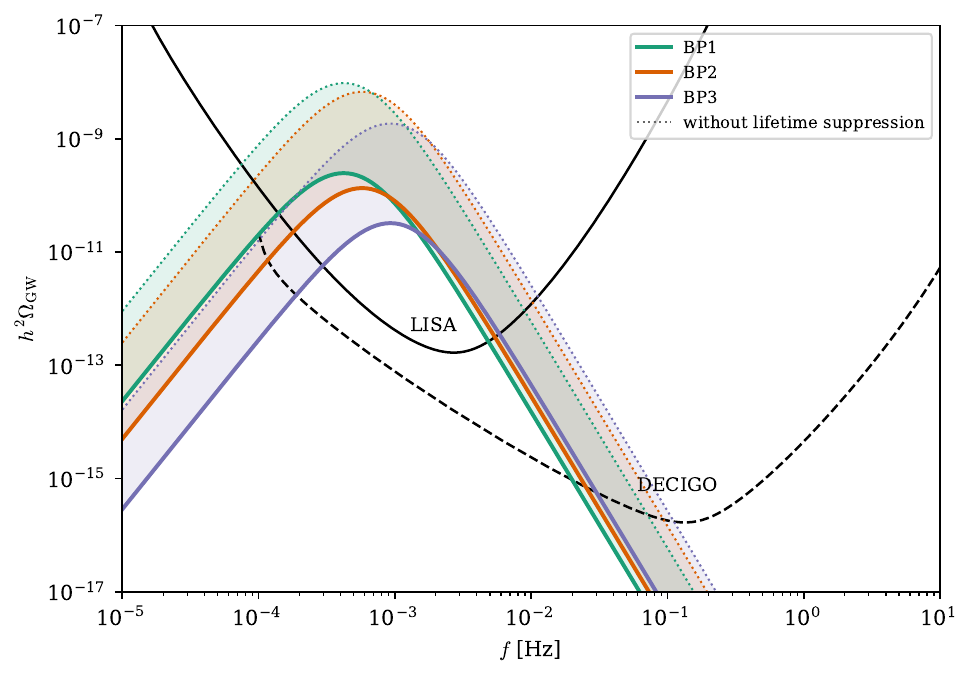}
\caption{Gravitational-wave spectra from sound waves for BP1--BP3. Solid curves include the finite lifetime of the source, eq.~\eqref{eq:Ups}, and dotted curves show the same spectrum with $\Upsilon=1$; the shaded bands lie between the two. Black curves are the power-law-integrated sensitivity curves of LISA (solid) and DECIGO (dashed) for four years of observation and SNR $=10$.}
\label{fig:spectra}
\end{figure}

Figure~\ref{fig:spectra} shows the spectra of the three benchmarks against the power-law-integrated sensitivity curves~\cite{Thrane:2013oya} of LISA~\cite{LISA:2017pwj,Robson:2018ifk} and DECIGO~\cite{Seto:2001qf,Yagi:2011wg} for four years of observation and a signal-to-noise threshold of ten. The peaks lie at $f_p\simeq0.42$, $0.58$ and $0.93$~mHz with amplitudes $h^2\Omega_{\rm sw}\simeq2.5\times10^{-10}$, $1.3\times10^{-10}$ and $3.2\times10^{-11}$. The peak frequencies fall just below the most sensitive part of the LISA band, and the spectra exceed the sensitivity curve over almost two decades in frequency. Since $f_p\propto\beta/H$ and the peak amplitude scales approximately as $(\beta/H)^{-2}$, the $20\%$ uncertainty on $\beta/H$ (section~\ref{sec:alphabeta}) translates into $20\%$ in $f_p$ and about $40\%$ in amplitude, a shift along the scaling relation of section~\ref{sec:universal}.

The lifetime factor is small, $\Upsilon\simeq0.026$, $0.020$ and $0.017$, because $\beta/H\sim10^2$ makes the bubbles small compared with the Hubble radius and the fluid velocity $\bar U_f\simeq0.7$--$0.86$ is large, so nonlinearities develop within a few percent of a Hubble time. Without this factor the amplitudes would be larger by a factor of $40$ to $60$ (dotted curves in figure~\ref{fig:spectra}). The suppression is the largest single correction in the estimate, and the shaded bands indicate how much depends on it.

We compute the signal-to-noise ratio as ${\rm SNR}^2=T_{\rm obs}\int df\,[\Omega_{\rm sw}(f)/\Omega_n(f)]^2$ with $T_{\rm obs}=4$~years, integrated from $10^{-6}$ to $10$~Hz, using the LISA noise model of ref.~\cite{Robson:2018ifk} without the galactic confusion foreground. Including that foreground would lower the LISA values somewhat, since the peaks lie near $1$~mHz. The ratios are about $1400$, $1300$ and $680$ for BP1, BP2 and BP3. DECIGO, whose sensitivity is greatest at higher frequencies, would observe the high-frequency tail of all three spectra.

%----------------------------------------------------------------------
\subsection{Scaling of the peak}
\label{sec:universal}
%----------------------------------------------------------------------

\begin{figure}[t]
\centering
\includegraphics[width=0.78\textwidth]{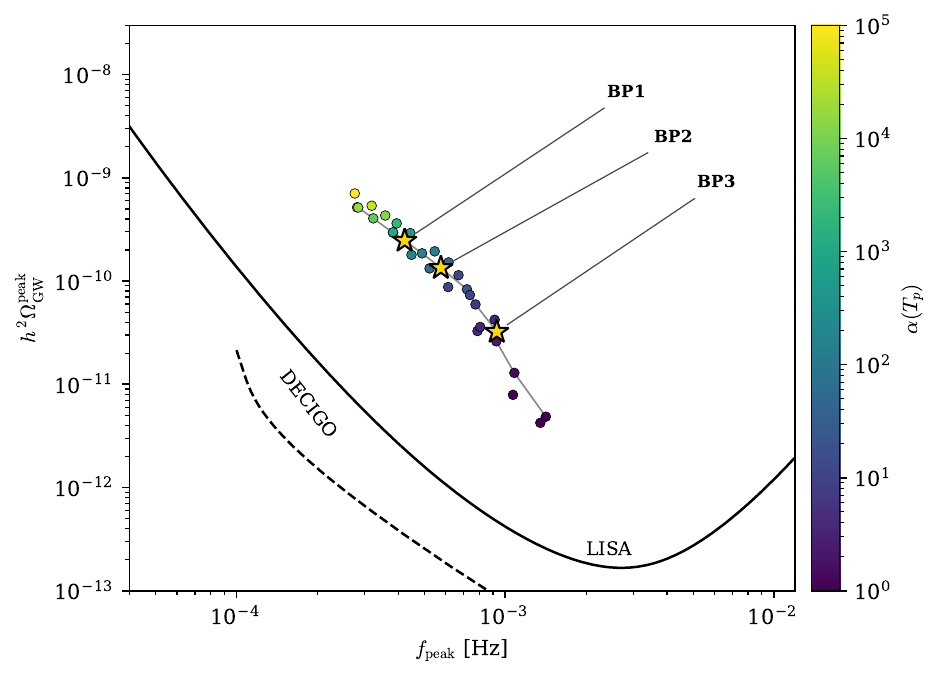}
\caption{Peak frequency and peak amplitude of the sound-wave spectrum, including the lifetime factor, for all computed points that satisfy the Higgs signal-strength, LEP, Landau-pole and perturbativity constraints, coloured by $\alpha(T_p)$. The points lie on five lines of fixed $\beta_Q/4\pi$ between $0.002$ and $0.02$; the line $\beta_Q/4\pi=0.006$ is connected. Stars mark BP1--BP3. Only points with $T_p\geq2$~GeV are shown. The black curves are the sensitivity curves of figure~\ref{fig:spectra}.}
\label{fig:peaks}
\end{figure}

Figure~\ref{fig:peaks} collects the peak frequency and amplitude of every computed point with $T_p\geq2$~GeV that satisfies the constraints of section~\ref{sec:constraints}. Despite the range of more than five orders of magnitude in $\alpha$, the points cluster in a narrow band. The reason follows from eqs.~\eqref{eq:Osw}--\eqref{eq:Ups}. For $\tau_{\rm sw}H_*\ll1$ the lifetime factor reduces to $\Upsilon\simeq R_*H_*/\bar U_f$, and eliminating $R_*H_*$ in favour of $f_p$ gives the approximate scaling relation
\begin{equation}
h^2\Omega_{\rm sw}^{\rm peak}\,f_p^2 \simeq 2.8\times10^{-16}\,{\rm Hz}^2\;\frac{K^2}{\bar U_f}\left(\frac{T_{\rm reh}}{100~{\rm GeV}}\right)^2 ,
\label{eq:universal}
\end{equation}
where we used $S(1)=1$. In a strongly supercooled transition $K\to1$ and $\bar U_f\to\sqrt3/2$, so the right-hand side depends only on $T_{\rm reh}$, which is set by $\Delta V_0^{1/4}$ and varies by less than a factor of two over the viable region. The inverse duration $\beta/H$ then moves a point along the line $h^2\Omega^{\rm peak}\propto f_p^{-2}$, but not off it. Equation~\eqref{eq:universal} reproduces the benchmark values to within a few percent, and the residual spread in figure~\ref{fig:peaks} comes from the variation of $T_{\rm reh}$ with $\beta_Q$ and from $K<1$ at the weakest points.

This makes the signal a fairly direct probe of the vacuum energy. Within the model, a measured peak frequency fixes $\beta/H$, and the amplitude at that frequency fixes $T_{\rm reh}$ and hence $\Delta V_0$. Because $\Delta V_0$ is generated by the same loops that give the dilaton its mass~\cite{Gildener:1976ih}, the amplitude is tied to the scalar spectrum of the model rather than to details of the transition dynamics.

%----------------------------------------------------------------------
\subsection{Detectability across the parameter space}
\label{sec:snr}
%----------------------------------------------------------------------

\begin{figure}[t]
\centering
\includegraphics[width=\textwidth]{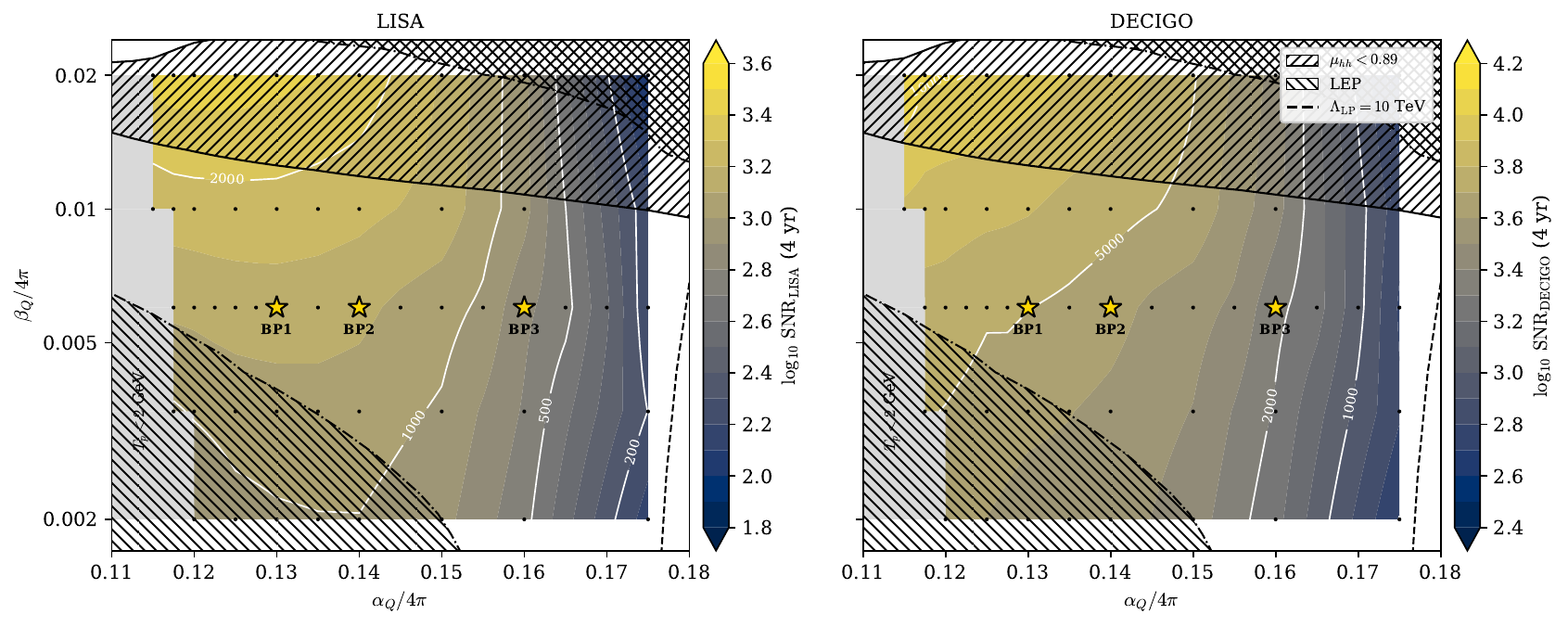}
\caption{Signal-to-noise ratio of the sound-wave signal for four years of observation with LISA (left) and DECIGO (right), in the $(\alpha_Q,\beta_Q)$ plane for $N_Q=12$ and $v_S=1$~TeV. The grey region has $T_p<2$~GeV, the region hatched to the right has $\mu_{hh}<0.89$, the region hatched to the left and bounded by the dash-dotted line is excluded by LEP, and the dashed line marks $\Lambda_{\rm LP}=10$~TeV. Dots are the computed points and stars mark BP1--BP3.}
\label{fig:snr}
\end{figure}

Figure~\ref{fig:snr} shows the signal-to-noise ratio over the $(\alpha_Q,\beta_Q)$ plane. Within the acoustic description, every viable point with $T_p\geq2$~GeV would be detected with high significance, with ${\rm SNR}_{\rm LISA}$ between $140$ and $1.8\times10^3$ and ${\rm SNR}_{\rm DECIGO}$ between $500$ and $7.6\times10^3$. The signal weakens toward large $\alpha_Q$, where the transition becomes faster and less supercooled, but even at the Landau-pole boundary it remains far above threshold. The LISA ratio is largest at intermediate $\alpha_Q$ and larger $\beta_Q$, where the peak frequency is highest among the strongly supercooled points and closest to the minimum of the LISA noise curve.

We exclude points with $T_p<2$~GeV from the gravitational-wave figures. In this range the bounce solutions show additional scatter, and the fitted $S_3/T$, whose derivative sets $\beta/H$, is less reliable than at higher temperatures. These points lie in the region where the spectrum already saturates eq.~\eqref{eq:universal}, so their omission does not affect the conclusions.

%======================================================================
\section{Dependence on $N_Q$ and $v_S$}
\label{sec:NQvS}
%======================================================================

The results so far were obtained for $N_Q=12$ and $v_S=1$~TeV. Neither number is fixed by the model. The multiplicity $N_Q$ is a free parameter of the hidden sector, and the singlet scale $v_S$ is bounded from below by Higgs measurements and from above only by naturalness considerations. This section shows how the conclusions change when these two parameters vary.

%----------------------------------------------------------------------
\subsection{Multiplicity of the singlets}
\label{sec:NQ}
%----------------------------------------------------------------------

The $Q_i$ affect the transition in two ways. At zero temperature they enter the one-loop potential through $N_Qm_Q^4$, which sets the dilaton mass, eq.~\eqref{eq:mdilaton}, and the depth of the vacuum. Along the Higgs direction this contribution is proportional to $N_Q\alpha_Q^2$. At finite temperature they contribute $N_Q\alpha_Q/24$ to the Higgs curvature $c_H$ of eq.~\eqref{eq:curv}. The first effect dominates, as the near collapse of the curves below confirms. The combination $N_Q\alpha_Q^2$ is therefore the leading scaling variable but not an exact one, and we compare different multiplicities in terms of $\alpha_Q\sqrt{N_Q/12}$, which reduces to $\alpha_Q$ for the reference choice. We keep $\beta_Q/4\pi=0.006$ fixed.

\begin{figure}[t]
\centering
\includegraphics[width=\textwidth]{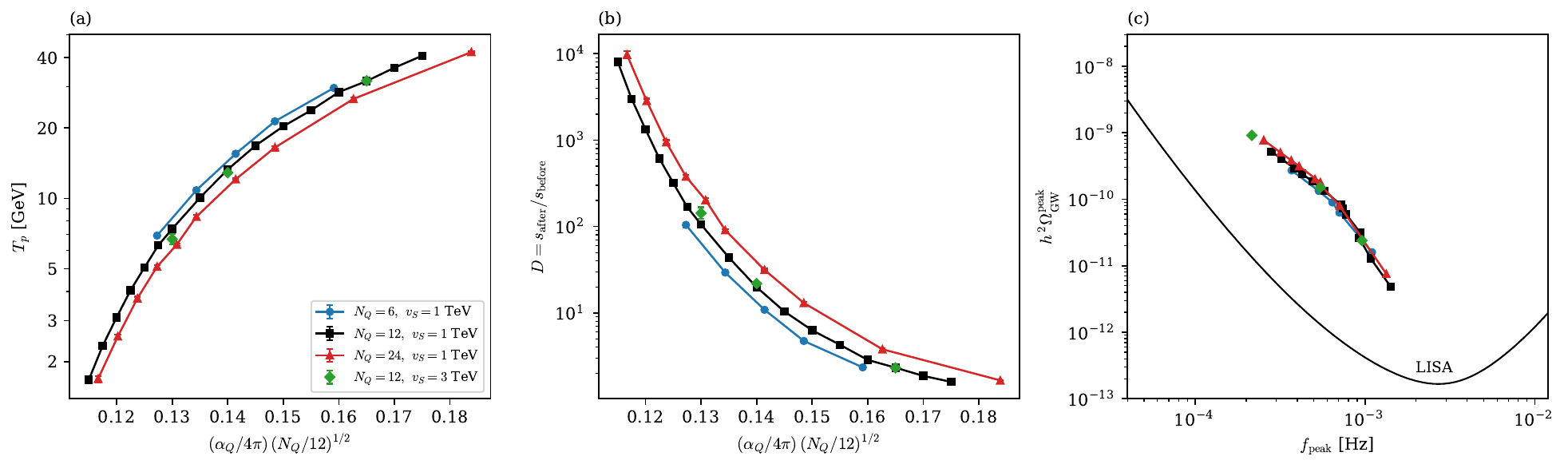}
\caption{Dependence on $N_Q$ and $v_S$. (a) Percolation temperature and (b) entropy injection against the scaled coupling $(\alpha_Q/4\pi)(N_Q/12)^{1/2}$, for $N_Q=6$, $12$ and $24$ at $v_S=1$~TeV and $\beta_Q/4\pi=0.006$, and for $N_Q=12$ at $v_S=3$~TeV with $\beta_Q$ rescaled to keep $\beta_Qv_S^2$ fixed. (c) Peak frequency and amplitude of the gravitational-wave spectrum, with the LISA sensitivity curve of figure~\ref{fig:spectra}. Only points satisfying $\mu_{hh}\geq0.89$, the LEP bound and $\Lambda_{\rm LP}>10$~TeV are shown, and in panel (c) only those with $T_p\geq2$~GeV. Error bars are the bootstrap intervals.}
\label{fig:NQvS}
\end{figure}

Figure~\ref{fig:NQvS} compares $N_Q=6$, $12$ and $24$ along this line. In terms of the scaled coupling, the percolation temperatures agree to within about $30\%$ (panel a), and the completion edge lies at $\alpha_Q\sqrt{N_Q/12}/4\pi\simeq0.113$--$0.117$ for all three multiplicities. The residual ordering, with lower $T_p$ for larger $N_Q$ at fixed $N_Q\alpha_Q^2$, comes from the thermal term. At fixed $N_Q\alpha_Q^2$ the contribution $N_Q\alpha_Q/24$ to $c_H$ grows as $\sqrt{N_Q}$, so the thermal barrier is larger and nucleation is delayed.

The entropy injection follows the same trend with a larger spread (panel b). Near the completion edge $D$ reaches about $8\times10^3$ for $N_Q=12$ and $10^4$ for $N_Q=24$. For $N_Q=6$ the dilaton is lighter and more strongly mixed, and the four points closest to the edge, with $D\simeq2\times10^2$--$2\times10^3$, are excluded by LEP; the largest allowed value is $D\simeq10^2$. Besides the lower $T_p$, the vacuum energy grows with $N_Q$ at fixed scaled coupling, because the contribution of $\beta_Q$ to $N_Qm_Q^4$ is not rescaled: at the edge $\Delta V_0^{1/4}$ increases from $80$ to $95$~GeV between $N_Q=6$ and $24$. Since $D\propto\Delta V_0^{3/4}/T_p^3$ for $\alpha\gg1$, both effects add up.

The gravitational-wave peaks of all three multiplicities follow the same scaling relation, eq.~\eqref{eq:universal} (panel c). The reheating temperature varies only between $34$ and $55$~GeV over all points shown, and a change of $N_Q$ only moves a point along the line $h^2\Omega^{\rm peak}\propto f_p^{-2}$.

The constraints scale less simply. The Higgs signal strength stays above $0.91$ for all points in figure~\ref{fig:NQvS}. The Landau pole moves to lower scales as the scaled coupling grows, and the condition $\Lambda_{\rm LP}>10$~TeV ends each curve at $\alpha_Q\sqrt{N_Q/12}/4\pi\simeq0.17$ for $N_Q=6$, $0.177$ for $N_Q=12$ and above $0.18$ for $N_Q=24$. Near the completion edge the Landau pole lies at $80$, $150$ and $190$~TeV. The strongly supercooled region is therefore compatible with perturbativity up to $\mathcal{O}(100)$~TeV for any multiplicity in this range.

%----------------------------------------------------------------------
\subsection{Singlet scale}
\label{sec:vS}
%----------------------------------------------------------------------

The singlet expectation value enters the transition through the mass $m_Q^2=(\alpha_Qv^2+\beta_Qv_S^2)/2$ and through the mixing of $h$ with the singlet. To isolate the second effect we compare $v_S=1$ and $3$~TeV at fixed $\beta_Qv_S^2$, that is with $\beta_Q/4\pi=0.006\times(1~{\rm TeV}/v_S)^2$, so that $m_Q$ and $\Delta V_0$ are the same. The resulting transitions are nearly identical. At $\alpha_Q/4\pi=0.13$, $0.14$ and $0.165$ we find $T_p\simeq6.9$, $12.9$ and $31.9$~GeV for $v_S=3$~TeV, compared with $7.4$, $13.3$ and $31.9$~GeV for $v_S=1$~TeV, the entropy injection agrees within $30\%$, and the gravitational-wave peaks coincide (figure~\ref{fig:NQvS}). This is expected from the tunnelling path of section~\ref{sec:path}. The bounce ends at $|s|$ of order $10$~GeV, far below $v_S$, and the singlet enters the dynamics only through the masses it generates. For $v_S=3$~TeV the field space is twelve times longer in the singlet direction than in the Higgs direction, and near the completion edge the two-dimensional bounce could not be determined reliably. Figure~\ref{fig:NQvS} therefore shows $v_S=3$~TeV only for $\alpha_Q/4\pi\geq0.13$. At $v_S=3$~TeV the dilaton mass drops to about $9$~GeV, below the range of the LEP bound of section~\ref{sec:constraints}, and $\sin^2\theta\leq2\times10^{-3}$; the constraints that apply at this mass have not been imposed.

The singlet scale is bounded from below by the Higgs signal strength. The mixing angle grows as $v_S$ decreases, and at $v_S=0.5$~TeV it gives $\mu_{hh}\simeq0.81$--$0.86$ across the range $0.10\leq\alpha_Q/4\pi\leq0.17$, which is excluded. At $v_S=0.7$~TeV the signal strength lies between $0.89$ and $0.93$, at the edge of the allowed range, while at $v_S=1$ and $3$~TeV it exceeds $0.95$ and $0.99$. Strong supercooling with sizeable entropy injection is therefore a feature of the model for $v_S\gtrsim1$~TeV, and above this value the singlet scale has little influence on the transition once $m_Q$ is held fixed. This last statement rests on the comparison at $\alpha_Q/4\pi\geq0.13$; closer to the completion edge we could not test it for $v_S=3$~TeV.
 
%======================================================================
\section{Conclusions}
\label{sec:conclusions}
%======================================================================

We have studied the electroweak phase transition in a classically scale-invariant extension of the Standard Model with a real singlet $s$ and $N_Q$ real scalars $Q_i$ that couple to the Higgs doublet and to the singlet through the portals $\alpha_Q$ and $\beta_Q$. Electroweak symmetry is broken radiatively along a Gildener-Weinberg flat direction, which leaves a light dilaton with $m_\eta\simeq27$~GeV, and the $Q_i$ supply the positive contribution to the one-loop potential that makes the vacuum stable.

The absence of tree-level mass terms keeps the symmetric phase metastable at all temperatures, and in the region where the $Q_i$ drive the transition it proceeds as a single, strongly supercooled step from the origin to the electroweak vacuum. We followed this step through nucleation, percolation and completion with the bounce action evaluated along the two-dimensional tunnelling path, and we draw the following conclusions.

\begin{enumerate}
\item The transition completes. Along the benchmark line the percolation temperature ranges from $1.7$ to $41$~GeV, the false-vacuum volume shrinks once bubbles percolate, and completion follows percolation within a few percent in temperature. The lower edge of the viable region is set by the onset of QCD effects at $T_p\simeq1.5$~GeV, not by a failure of percolation. The tunnelling path leaves the origin along the Higgs direction and turns into the flat direction only during the classical roll inside the bubble, and a straight-line path overestimates the bounce action by an order of magnitude.

\item Despite strong supercooling the transition is fast, with $\beta/H\simeq60$--$320$. The bounce action is still decreasing at $T_p$, and its logarithmic slope multiplied by $S_3/T\simeq135$ gives values of order $10^2$. Small $\beta/H$, often associated with supercooling, does not occur here.

\item Nearly all of the released energy goes into the plasma. For $\alpha_Q/4\pi\lesssim0.14$ at $N_Q=12$ the walls reach terminal Lorentz factors between several tens and a few times $10^4$, set by next-to-leading-order friction, long before bubbles collide. For somewhat weaker transitions, including BP3, the $Q_i$ are reflected by the wall and a simple estimate still gives Lorentz factors of a few. In both cases the fraction of energy in the scalar field is negligible, and the kinetic fraction of the fluid approaches one for $\alpha\gg1$.

\item The transition injects entropy. Reheating takes the plasma to $T_{\rm reh}\simeq30$--$55$~GeV, close to the temperature at which vacuum energy and radiation would be equal, while the universe had supercooled to $T_p$. The resulting entropy injection grows from $D\simeq1.5$ at the weak end of the viable region to $D\gtrsim10^3$ near the completion edge, and $D\gtrsim100$ wherever $\alpha_Q/4\pi\lesssim0.13$ at $N_Q=12$. Pre-existing relics are diluted accordingly. A baryon asymmetry produced earlier is reprocessed by sphalerons in the supercooled symmetric phase and then diluted, so that the required initial $B-L$ asymmetry is larger than the observed baryon asymmetry by $(79/28)D$. A thermal relic that freezes out before the transition can be as heavy as $m_U\sqrt D$, of order $10^6$--$10^7$~GeV for the strongest transitions, and any earlier gravitational-wave background is suppressed by $D^{4/3}$ relative to radiation. These statements concern species decoupled before the transition. The $Q_i$ are not among them: they re-equilibrate after reheating and decay before nucleosynthesis through small couplings that break their $O(N_Q)$ symmetry.

\item If the acoustic description applies, the gravitational-wave signal is large. Its peak lies between $0.3$ and $1.5$~mHz with $h^2\Omega_{\rm peak}$ between a few times $10^{-12}$ and $10^{-9}$, and the product $h^2\Omega_{\rm peak}f_p^2$ follows an approximate scaling relation that depends mainly on $T_{\rm reh}$, which is fixed by the vacuum energy. The transition rate $\beta/H$ moves a point along this relation in the plane of peak frequency and amplitude, and the computed points with $T_p\geq2$~GeV give signal-to-noise ratios between $140$ and about $2\times10^3$ in four years of LISA observation.

\item The results depend on the multiplicity and the portal at leading order through the combination $N_Q\alpha_Q^2$, with a residual dependence on $N_Q$ at the level of tens of percent in $T_p$ and a factor of a few in $D$. They are insensitive to the singlet scale at fixed $m_Q$, while the Higgs signal strength requires $v_S\gtrsim1$~TeV.
\end{enumerate}

The gravitational-wave amplitudes carry the largest uncertainty of the analysis. The spectra rely on a description of acoustic production that has been calibrated on simulations with $\alpha\lesssim0.1$, while the transitions considered here reach $\alpha\sim10^5$. Simulations of stronger transitions show that less energy ends up in long-lived sound waves, and for $\alpha\gg1$ the relativistic fluid shells may source a spectrum closer to that of colliding thin walls. A reduction of the amplitude by an order of magnitude would still leave the computed points above the LISA threshold, because the signal-to-noise ratios exceed it by factors of $14$ to $180$. The peak frequency, which is fixed by $\beta/H$ and $T_{\rm reh}$, is the more robust prediction. Among the inputs to the transition itself, the resummation prescription for the thermal potential is the dominant source of uncertainty, followed by the renormalisation scale of the thermal potential, which changes $T_p$ by up to seven percent and $\alpha$ by up to thirty percent between $\mu=\pi T_p$ and $4\pi T_p$ (appendix~\ref{app:scheme}). These comparisons cover the benchmarks only and are a limited robustness check rather than an estimate of higher-order corrections, which would require a dimensionally reduced effective theory and a two-loop treatment.

The model can also be tested at colliders. The benchmark points have a dilaton with $m_\eta\simeq27$~GeV, a branching ratio ${\rm BR}(h\to\eta\eta)\simeq2$--$3\%$, and Higgs signal strengths reduced by $2$--$4\%$ at $v_S=1$~TeV. The rare final state $b\bar b\mu^+\mu^-$ lies about an order of magnitude below current LHC limits, but the dominant decay into four $b$ quarks and the reduction of the signal strengths are both within the projected reach of a future lepton collider~\cite{Liu:2016zki}.

Several extensions suggest themselves. Below $T\simeq1.5$~GeV the transition would be triggered by QCD effects, and following it into that regime would extend the region of large entropy injection beyond the edge found here. A calculation of the wall velocity beyond the leading-order friction estimate is needed for $\alpha_Q/4\pi\gtrsim0.15$, including BP3, where the $Q_i$ are reflected by the wall. Most importantly, a dedicated numerical study of acoustic production for $\alpha\gg1$ with ultrarelativistic walls would turn the amplitude estimates of this work into firm predictions for the space-based interferometers.
 
\section{Acknowledgment}
This work was financially supported by Vellore Institute of Technology (VIT), Vellore under the Faculty Seed Grant (RGEMS) (Sanction Order No.: SG20260024).

\appendix
%======================================================================
\section{Numerical accuracy of the bounce action}
\label{app:numerics}
%======================================================================

\begin{figure}[t]
\centering
\includegraphics[width=\textwidth]{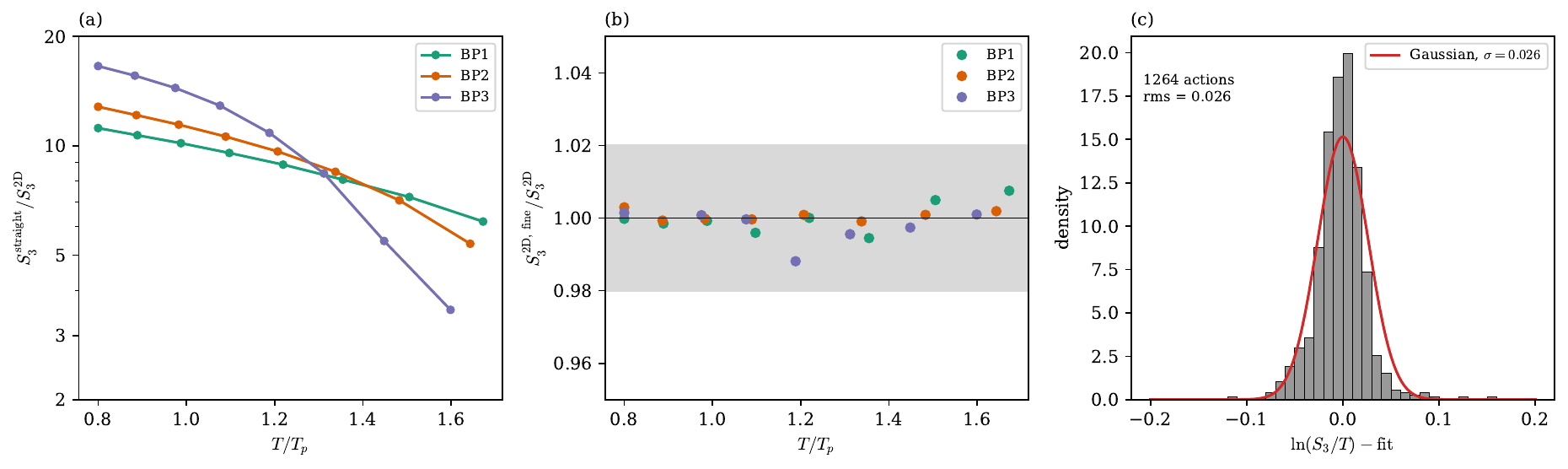}
\caption{Numerical checks of the bounce action for BP1--BP3. (a) Action along the straight path between the minima relative to the path-deformed action. (b) Path-deformed action with tighter solver settings relative to the default settings; the band marks $\pm2\%$. (c) Residuals of $\ln(S_3/T)$ about the smoothing fit for $1264$ bounce solutions near the transition, with a Gaussian of the measured rms width.}
\label{fig:numerics}
\end{figure}

Every result of sections~\ref{sec:perc}--\ref{sec:NQvS} rests on the bounce action $S_3(T)$ and on the fit used to differentiate and integrate it. This appendix documents how accurately both are determined for the one-loop, daisy-resummed potential of section~\ref{sec:VT}. Figure~\ref{fig:numerics} summarises the checks for the benchmark points. Numerical accuracy at the percent level does not imply the same accuracy for the physical predictions, which are limited by the theoretical uncertainties discussed in appendix~\ref{app:scheme}.

\paragraph{Path deformation.}
The $O(3)$ bounce is computed with the path-deformation method of \texttt{CosmoTransitions}~\cite{Wainwright:2011kj}, which iterates between a one-dimensional bounce along a trial path and a deformation of the path in the $(H,s)$ plane. The method converges to a critical path, which need not be the one of lowest action. Started from the straight line between the two minima, it can settle on a path that crosses the barrier far from the Higgs axis, with an action that is too large by up to a factor of two. We therefore solve each bounce twice, once from the straight line and once from a path that first leaves the false vacuum along $H$, and keep the lower action. At the percolation temperatures of the benchmarks, four further initial paths with different kinks change $S_3/T$ by less than one percent, so the lower of the two solutions is the minimum within the accuracy of the method. Both initial paths are used for every bounce in the scans of sections~\ref{sec:history}--\ref{sec:NQvS}, not only at the benchmarks. If path deformation fails to return a finite action, the straight-path action is recorded instead. Since it lies far above the path-deformed curve, such a point is removed by the outlier criteria below, and the straight-path action never replaces a finite path-deformed solution.

For $v_S=3$~TeV the distance between the minima is twelve times longer in the singlet direction than in the Higgs direction, and two additional initial paths, with kinks scaled to the Higgs direction, are included. Near the completion edge this is still not sufficient: $S_3/T$ jumps between neighbouring temperatures, and these points are omitted from figure~\ref{fig:NQvS}.

\paragraph{Comparison with the straight path.}
The action along the straight line between the minima is an upper bound on the true bounce action. Figure~\ref{fig:numerics}a shows that it exceeds the path-deformed result by factors of $3.5$ to $17$ between $0.8\,T_p$ and $1.7\,T_p$. The ratio is largest at low temperature, where the barrier is confined to field values of a few times $T$ near the Higgs axis and the straight path crosses it where it is widest. A single-field treatment along the flat direction would therefore place nucleation at far too low a temperature, or exclude completion altogether.

\paragraph{Solver settings.}
Figure~\ref{fig:numerics}b compares the default settings with a run that allows $30$ instead of $10$ deformation iterations, samples the potential along the path at $300$ points, and tightens the tolerances of the one-dimensional profile. The two agree with an rms deviation of $0.4\%$ and a largest deviation of $1.2\%$ over $23$ temperatures. At one temperature of BP3, $T=25.07$~GeV or $0.88\,T_p$, the tighter run did not converge, and this temperature is omitted from the comparison.

\paragraph{Fit and bootstrap.}
The bounce action is computed on a grid of $26$ temperatures between $0.97\,T_c$ and $1.5$~GeV, refined by $11$ temperatures between $0.8\,T_p$ and $1.3\,T_n$. Individual solutions scatter around a smooth curve, and $S_3/T$ is represented by a cubic smoothing spline in $(\ln T,\ln S_3/T)$ with smoothing condition $\sum_i r_i^2\leq N\sigma^2$, where $r_i$ are the residuals and $N$ the number of points. Figure~\ref{fig:numerics}c shows the distribution of residuals for $1264$ bounce solutions near the transition, collected from the benchmarks and from all lines of section~\ref{sec:Dplane} with $T_p\geq2$~GeV. Its core is Gaussian with an rms of $0.026$, and about $1.4\%$ of the solutions lie in a tail beyond $0.08$. Throughout the paper we use $\sigma=0.04$, which accommodates this tail. Since $S_3/T$ is continuous in $T$, an isolated solution that deviates from the interpolation between its two neighbours by more than $0.15$ in $\ln(S_3/T)$, and by more than the difference between the neighbours themselves, is treated as a failed solution and discarded. A solution that lies more than $25\%$ above the median of its neighbours is also discarded, because the bounce action is the minimum over paths and an upward excursion signals a deformation that settled on the wrong critical path. Near the transition this criterion removes two solutions at $v_S=3$~TeV and one in the scheme comparison of appendix~\ref{app:scheme}, and none elsewhere. The uncertainties quoted in tables~\ref{tab:transition} and~\ref{tab:budget} and shown as error bars in the figures come from a residual bootstrap: the residuals are resampled with replacement, added to the fitted curve, refitted, and the percolation analysis is repeated, with $300$ samples for the benchmarks and $80$ to $120$ for the other points. We quote the median and the $16$th and $84$th percentiles. The residual bootstrap treats the residuals as independent, whereas neighbouring residuals are positively correlated, with a mean lag-one correlation of $0.7$. A block bootstrap that resamples blocks of three or five consecutive residuals gives intervals for $T_p$, $\beta/H$ and $D$ at the benchmarks that are no wider than those of the ordinary bootstrap, so the correlation does not enlarge the quoted uncertainties.

\paragraph{Dependence on the smoothing.}
The bootstrap does not capture the dependence on the assumed $\sigma$ itself. Table~\ref{tab:sigma} shows the effect of replacing $\sigma=0.04$ by the measured rms or by a smaller value that lets the fit follow individual solutions more closely. The percolation temperature changes by at most $1\%$ at the benchmarks and $5\%$ at the completion edge, and the entropy injection by at most $3\%$ and $17\%$, respectively. The inverse duration is more sensitive, because it depends on the derivative of the fit, and changes by up to $20\%$. We regard $20\%$ as the numerical uncertainty of $\beta/H$, compared with about $5\%$ suggested by the bootstrap alone. This is still small compared with the uncertainty of the gravitational-wave amplitude discussed in section~\ref{sec:conclusions}, and it does not affect any of the conclusions.

\paragraph{Dependence on the temperature grid.}
The refinement temperatures between $0.8\,T_p$ and $1.3\,T_n$ are placed according to a first estimate of $T_p$ and $T_n$, so two runs of the same potential sample $S_3/T$ at different temperatures. To isolate this effect we use two runs whose actions agree to better than $0.1\%$ at common temperatures (those with the two interpolations of the thermal functions described below) and fit each grid alone and their union. Across these fits $T_p$ varies by up to $2\%$, $\alpha$ by up to $7\%$ and $D$ by up to $5\%$, with the largest spread at BP2, and $\beta/H$ by up to $13\%$. The shifts in $T_p$, $\alpha$ and $D$ exceed the bootstrap intervals, and we regard them as an additional numerical systematic of the benchmark values; the shift in $\beta/H$ is covered by the $20\%$ assigned above.

\paragraph{Interpolation of the thermal functions.}
The functions $J_B$ and $J_F$ of eq.~\eqref{eq:JBJF} are tabulated and interpolated with cubic splines in $y=m^2/T^2$. Because $J_B$ contains a term proportional to $y^{3/2}$, a spline in $y$ underestimates its slope at $y\to0$ by $4.7\%$, and with it the $T^2m^2/24$ terms of the bosons near the origin. A spline in $\sqrt{y}$ reproduces the slope. With this interpolation the bounce action between $0.7\,T_p$ and $1.5\,T_p$ changes by less than $0.1\%$ at BP1 and BP2 and by at most $1\%$ at BP3. The derived parameters shift by $+0.6\%$, $-1.7\%$ and less than $0.1\%$ in $T_p$, by $-2.4\%$, $+7.2\%$ and less than $0.1\%$ in $\alpha$, and by $-1.8\%$, $+5.3\%$ and less than $0.1\%$ in $D$ at BP1, BP2 and BP3, and $\beta/H$ and $T_{\rm reh}$ by at most $1.1\%$. Given the near equality of the actions, these shifts reflect the different refinement grids rather than the interpolation itself. All results in this paper use the spline in $y$.

\begin{table}[t]
\centering
\renewcommand{\arraystretch}{1.2}
\begin{tabular}{lccc|ccc|ccc}
\hline\hline
 & \multicolumn{3}{c|}{$T_p$ [GeV]} & \multicolumn{3}{c|}{$\beta/H$} & \multicolumn{3}{c}{$D$} \\
$\sigma$ & 0.015 & 0.026 & 0.04 & 0.015 & 0.026 & 0.04 & 0.015 & 0.026 & 0.04 \\ \hline
BP1 & 7.35 & 7.46 & 7.39 & 111 & 103 & 125 & 108 & 103 & 106 \\
BP2 & 13.22 & 13.29 & 13.31 & 143 & 139 & 156 & 20 & 20 & 20 \\
BP3 & 28.08 & 28.32 & 28.38 & 225 & 234 & 235 & 2.94 & 2.88 & 2.87 \\
$\alpha_Q/4\pi=0.115$ & 1.60 & 1.69 & 1.69 & 63 & 62 & 62 & 9211 & 7930 & 7868 \\
\hline\hline
\end{tabular}
\caption{Dependence of the percolation temperature, the inverse duration and the entropy injection on the width $\sigma$ assumed for the scatter of $\ln(S_3/T)$ in the smoothing fit. The value $\sigma=0.04$ is used throughout the paper, $\sigma=0.026$ is the measured rms residual, and $\sigma=0.015$ allows the fit to follow the individual bounce solutions more closely. The entries are single fits to the computed actions, whereas table~\ref{tab:transition} quotes bootstrap medians. The two differ by at most $3\%$ in $T_p$ and $D$, and by up to $8\%$ in $\beta/H$ at BP2, where the bootstrap distribution of the derivative is asymmetric. The last row is the point closest to the completion edge on the benchmark line.}
\label{tab:sigma}
\end{table}

%======================================================================
\section{Dependence on the thermal resummation}
\label{app:scheme}
%======================================================================

The finite-temperature potential of section~\ref{sec:VT} is a one-loop expression with the leading infrared contributions resummed. Different ways of performing that resummation agree at the order to which they are consistent and differ beyond it, and the size of the difference is a measure of the uncertainty from higher orders~\cite{Curtin:2016urg}. For a strongly supercooled transition this uncertainty is amplified, because the percolation temperature depends logarithmically on the nucleation rate while the strength $\alpha\propto T_p^{-4}$ responds to $T_p$ with a large power. This appendix compares the default treatment with three variations of the resummation and with a variation of the renormalisation scale, at the three benchmarks. The comparison is limited to these points and to variations within the one-loop framework, so it indicates the size of the scheme dependence rather than bounding the higher-order corrections.

\paragraph{Default.}
Arnold-Espinosa resummation~\cite{Arnold:1992rz}, eq.~\eqref{eq:daisy}, applied to the longitudinal gauge bosons and to the $Q_i$ with the Debye mass~\eqref{eq:PiQ}. Thermal loops of $h$, $\eta$ and the Goldstone modes are omitted.

\paragraph{Parwani resummation.}
The thermal masses are inserted into the full thermal functions, $J_B(m_i^2/T^2)\to J_B\big((m_i^2+\Pi_i)/T^2\big)$, for the same species, and the separate daisy term is dropped~\cite{Parwani:1991gq}. This resums the thermal masses also in the non-zero Matsubara modes, where they are formally of higher order.

\paragraph{Scalar loops.}
The default potential is supplemented by the thermal loops of the two neutral scalar mass eigenstates and the three Goldstone modes. Their field-dependent masses are obtained from the curvature of the zero-temperature effective potential. Where a squared mass is negative we set it to zero in $J_B$; this is a prescription that discards the imaginary part of the potential and is not derived from the resummed theory. The states are resummed with the Debye masses
\begin{equation}
\Pi_H = T^2\left[\frac{3g^2+g'^2}{16}+\frac{y_t^2}{4}+\frac{\lambda_H}{12}+\frac{\omega}{24}+\frac{N_Q\alpha_Q}{24}\right],
\qquad
\Pi_S = T^2\left[\frac{\omega}{6}+\frac{\lambda_S}{24}+\frac{N_Q\beta_Q}{24}\right],
\label{eq:PiHS}
\end{equation}
which are the diagonal entries of the thermal mass matrix in the $(H,s)$ basis. We add them to the mostly-doublet and mostly-singlet eigenstates, which neglects the off-diagonal thermal mixing; with $\sin^2\theta\leq0.013$ at the benchmarks this is a small effect. With scalar loops included, one default-setting bounce solution at BP1, at $T=8.64$~GeV, lies $30\%$ below its neighbours, with $S_3/T\simeq108$--$111$ for different initial paths. With the tighter solver settings of appendix~\ref{app:numerics} the same temperature gives $S_3/T\simeq155$--$157$ for three initial paths, in line with the neighbours, so the low value is a solver failure. Such solutions are removed by the criterion of appendix~\ref{app:numerics}.

\paragraph{Generated $\lambda_Q$.}
With $\lambda_Q(m_h)=0$, running to $\mu\simeq\pi T_p$--$2\pi T_p$ shifts $\Pi_Q$ by between $-22\%$ and $+6\%$ at the benchmarks (section~\ref{sec:tree}). We bracket this effect by lowering $\Pi_Q$ of eq.~\eqref{eq:PiQ} by $15\%$ at all three benchmarks, which is representative of BP1 and BP2 and larger than the shift expected at BP3.

\begin{figure}[t]
\centering
\includegraphics[width=\textwidth]{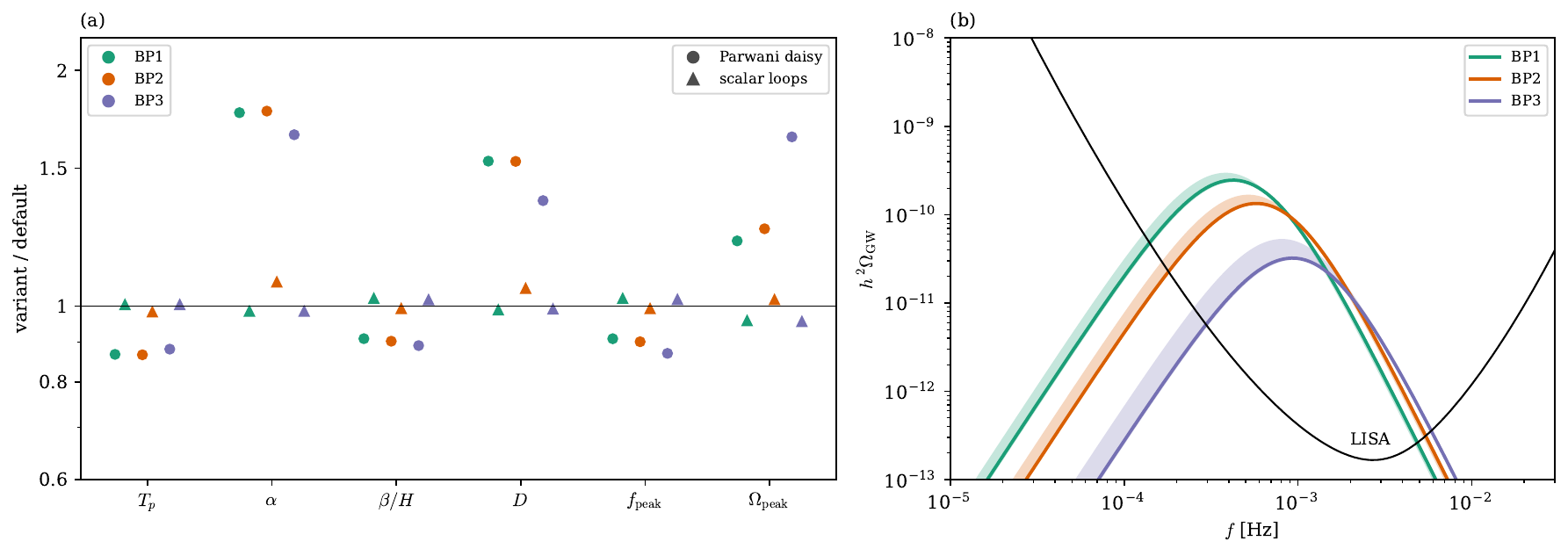}
\caption{Dependence on the thermal resummation for BP1--BP3. (a) Ratio of $T_p$, $\alpha$, $\beta/H$, $D$, and the peak frequency and amplitude of the gravitational-wave spectrum to their default values, for the Parwani prescription (circles) and with scalar thermal loops included (triangles). (b) Spectra of figure~\ref{fig:spectra} (lines) and the range spanned by the two variations (bands), with the LISA sensitivity curve.}
\label{fig:scheme}
\end{figure}

\paragraph{Results.}
Figure~\ref{fig:scheme}a shows the ratio of each quantity to its default value for the Parwani prescription and the scalar loops. The scalar loops have little effect. The bounce action at common temperatures changes by at most $2\%$, and the transition parameters by at most $7\%$, with the largest shift, at BP2, of the size of the grid systematic of appendix~\ref{app:numerics}. The effect is small because the scalar contributions to the thermal masses and to the barrier are small compared with those of the gauge bosons, the top quark and the $Q_i$. In $\Pi_H$, for example, the $Q_i$ contribute $N_Q\alpha_Q/24\simeq0.8$ at BP1 and the Higgs quartic only $\lambda_H/12\simeq0.03$. The Parwani prescription has a larger effect. It lowers the percolation temperature by $12$--$13\%$ at all three benchmarks, which increases the strength by a factor of $1.65$--$1.8$ and the entropy injection by a factor of $1.35$--$1.55$, and it lowers $\beta/H$ by about $10\%$. The reheating temperature is unchanged within $3\%$, since it is set by the vacuum energy. The peak frequency of the gravitational-wave spectrum decreases by $10$--$14\%$ and the peak amplitude increases by $20$--$60\%$.

Lowering $\Pi_Q$ by $15\%$ reduces the thermal barrier and acts in the same direction, but less strongly: $T_p$ decreases by $2$--$4\%$, $\alpha$ increases by $10$--$17\%$, $D$ by $7$--$12\%$, and $\beta/H$ decreases by $7$--$20\%$, while $T_{\rm reh}$ is unchanged within one percent. A thermal mass of the $Q_i$ that is uncertain at the ten-percent level is therefore a smaller effect than the choice of resummation prescription.

These shifts are similar across the benchmarks and do not change any qualitative conclusion. The completion edge moves to slightly larger $\alpha_Q$, since a lower $T_p$ at fixed couplings brings the edge points below $1.5$~GeV, but only by about $0.001$ in $\alpha_Q/4\pi$, because $T_p$ rises steeply with $\alpha_Q$ there (figure~\ref{fig:temps}). Figure~\ref{fig:scheme}b shows the spectra of figure~\ref{fig:spectra} together with the envelope of the Parwani and scalar-loop variations. The scheme dependence of the amplitude is well below the uncertainty from the gravitational-wave production discussed in section~\ref{sec:conclusions}, and the shift of the peak frequency is comparable to the numerical uncertainty of $\beta/H$ (appendix~\ref{app:numerics}). \paragraph{Renormalisation scale.}
The couplings of the thermal potential are evaluated at $\mu=m_h$ by default, while the relevant thermal scale is of order $\pi T_p$, which lies a factor of $1.4$ to $5$ below $m_h$ at the benchmarks. The zero-temperature one-loop potential is fixed by the physical conditions of section~\ref{sec:oneloop} and is independent of the renormalisation scale at this order, so we keep it unchanged. We run $g$, $g'$, $y_t$, $\alpha_Q$, $\beta_Q$ and $\lambda_Q$ at one loop from $m_h$, with $\lambda_Q(m_h)=0$, to $\mu=\pi T_p$, $2\pi T_p$ and $4\pi T_p$, and use the running values in the thermal functions, the field-dependent masses that enter them and the Debye masses, including the contribution of $\lambda_Q$ to $\Pi_Q$. Over this range $\alpha_Q$ changes by up to ten percent. Table~\ref{tab:mu} lists the results. The percolation temperature moves by between $-7\%$ and $+5\%$, the strength by between $-19\%$ and $+31\%$, the entropy injection by between $-11\%$ and $+22\%$, and $T_{\rm reh}$ by at most about one percent. The inverse duration changes by up to $24\%$, which is comparable to its numerical uncertainty; its non-monotonic dependence on $\mu$ at BP1 reflects the scatter of the fitted derivative rather than the scale dependence itself. The peak amplitude, which scales approximately as $(\beta/H)^{-2}$, inherits this scatter and changes by between $-21\%$ and $+72\%$. The scale dependence is therefore smaller than the dependence on the resummation prescription for $T_p$, $\alpha$ and $D$, and comparable to it for the gravitational-wave amplitude. Lower scales give stronger transitions at BP1 and BP2, where the default scale $m_h$ lies above the range considered, so the default values there sit at the weak end of the band.

\begin{table}[t]
\centering
\setlength{\tabcolsep}{5pt}
\begin{tabular}{l ccc ccc ccc}
\hline\hline
 & \multicolumn{3}{c}{BP1} & \multicolumn{3}{c}{BP2} & \multicolumn{3}{c}{BP3} \\
$\mu/\pi T_p$ & 1 & 2 & 4 & 1 & 2 & 4 & 1 & 2 & 4 \\
\hline
$T_p$ & 0.94 & 0.96 & 0.99 & 0.95 & 0.97 & 1.00 & 0.97 & 1.02 & 1.05 \\
$\alpha$ & 1.31 & 1.18 & 1.04 & 1.24 & 1.12 & 1.02 & 1.15 & 0.92 & 0.81 \\
$\beta/H$ & 0.81 & 0.76 & 0.90 & 0.86 & 0.92 & 1.01 & 0.98 & 1.04 & 1.05 \\
$D$ & 1.22 & 1.13 & 1.03 & 1.17 & 1.09 & 1.02 & 1.08 & 0.95 & 0.89 \\
$T_{\rm reh}$ & 1.00 & 1.00 & 1.00 & 1.00 & 1.00 & 1.00 & 0.99 & 1.01 & 1.01 \\
$h^2\Omega_{\rm peak}$ & 1.51 & 1.72 & 1.24 & 1.35 & 1.20 & 0.98 & 1.13 & 0.88 & 0.79 \\
\hline\hline
\end{tabular}
\caption{Dependence on the renormalisation scale of the thermal potential. Each entry is the ratio to the default value, for couplings in the thermal functions and Debye masses run to $\mu=\pi T_p$, $2\pi T_p$ and $4\pi T_p$, with $T_p$ the default percolation temperature. The default calculation corresponds to $\mu=m_h$, that is $\mu/\pi T_p=5.4$, $3.0$ and $1.4$ at BP1, BP2 and BP3. The peak frequency changes in proportion to $\beta/H$.}
\label{tab:mu}
\end{table}

A dimensionally reduced effective theory at higher order~\cite{Kajantie:1995dw,Ekstedt:2022bff} would reduce this uncertainty, but its application to a supercooled transition, where $T_p$ lies far below the masses of the heavy states, requires a separate treatment of the heavy and light sectors that goes beyond the scope of this work.

\bibliography{ref}
\end{document}